\documentclass{rsproca_new}

\usepackage{graphicx}
\usepackage{epstopdf, epsfig}

\usepackage{comment}

\usepackage{graphicx}
\usepackage{amsmath}%
\usepackage{longtable}%
\usepackage{bm}%
\usepackage{color}
\usepackage[normalem]{ulem}

\definecolor{mycol1}{rgb}{0.9047,0.1918,0.1988}
\definecolor{mycol2}{rgb}{0.2941,0.5447,0.7494}
\definecolor{mycol3}{rgb}{0.3718,0.7176,0.3612}
\definecolor{mycol4}{rgb}{1.0000,0.5482,0.1000}
\definecolor{mycol5}{rgb}{0.8650,0.8110,0.4330}
\definecolor{mycol6}{rgb}{0.6859,0.4035,0.2412}
\definecolor{mycol7}{rgb}{0.9718,0.5553,0.7741}
\definecolor{mycol8}{rgb}{0.6400,0.6400,0.6400}
\definecolor{mycol9}{rgb}{0.6365,0.3753,0.6753}

\newcommand{\rfig}[1]{figure~\ref{fig:#1}}
\newcommand{\req}[1]{(\ref{eq:#1})}

\newcommand{\colr}[1]{\textcolor{red}{#1}}

\DeclareUnicodeCharacter{2212}{-}

\begin{document}

%%%% Article title to be placed here
\title{Laws of turbulence decay from direct numerical simulations}

\author{%%%% Author details
John Panickacheril John$^{1}$, Diego A Donzis$^{2}$ and Katepalli R Sreenivasan $^{3}$}

%%%%%%%%% Insert author address here
\address{$^{1}$Department of Mechanical and Aerospace Engineering,  New York University, New York, NY 10012, USA \\
$^{2}$Department of Aerospace Engineering, Texas A\&M University, College Station, TX 77843, USA \\
$^{3}$Department of Mechanical and Aerospace Engineering, Department of Physics and Courant Institute of Mathematical Sciences, New York University, New York, NY 10012, USA}

%%%% Subject entries to be placed here %%%%
\subject{Decaying Turbulence}

%%%% Keyword entries to be placed here %%%%

%%%% Insert corresponding author and its email address}
\corres{K.R. Sreenivasan \\
\email{katepalli.sreenivasan@nyu.edu}}

%%%% Abstract text to be placed here %%%%%%%%%%%%
\begin{abstract}
Inspection of available data on the decay exponent for the kinetic energy of homogeneous and isotropic turbulence (HIT) shows that it varies by as much as 100\%. Measurements and simulations often show no correspondence with theoretical arguments, which are themselves varied. This situation is unsatisfactory
given that HIT is a building block of turbulence theory and modeling. We take recourse to a large base of direct numerical simulations and study decaying HIT for a variety of initial conditions. We show that the Kolmogorov decay exponent and the Birkhoff-Saffman decay are both readily observed, albeit approximately, for long periods of time if the initial conditions are appropriately arranged. We also present, for both cases, other turbulent statistics such as the velocity derivative skewness, energy spectra and dissipation, and show that the decay and growth laws are approximately as expected theoretically, though the wavenumber spectrum near the origin begins to change relatively quickly, suggesting that the invariants do not strictly exist. We comment briefly on why the decay exponent has varied so widely in past experiments and simulations. 
\end{abstract}
%%%%%%%%%%%%%%%%%%%%%%%%%%%
\maketitle
\section{Introduction} 
Homogeneous and isotropic turbulence (HIT) has inspired many theoretical studies of turbulence (see, e.g., Refs. \cite{batch1953,pope2000,mccombbook2014}), and provided the basic modeling constant on energy dissipation
\cite{HLbook2011}. It is the simplest turbulent flow possible and its conceptually minimalistic dynamics consists of a monotonic decay of turbulent kinetic energy, with a resulting rearrangement of the energy spectral density. HIT has been studied for more than sixty years but remains somewhat murky even today with basic contradictions and unresolved anomalies. Our goal is to address them by using large direct numerical simulations (DNS) as the tool. This approach allows us a degree of control on initial conditions that is not attained in grid turbulence experiments.

Once HIT is produced in some manner (experimentally by sweeping a grid of bars or by pushing a flow past the grid, or, computationally by injecting initial energy into some chosen wave number bands), the equation for the average turbulent kinetic energy reduces to \cite{batch1953}
\begin{equation}
\frac{dk}{dt} = - \langle \epsilon \rangle,  
\label{eq:kdecay}
\end{equation}
where the turbulent kinetic energy $k = \frac{3}{2} u^{\prime 2}$, 
$u' = \sqrt{\langle u_{i} u_{i} \rangle /3}$ 
being the root-mean-square velocity, and the average rate of turbulent energy dissipation 
$\langle \epsilon \rangle \equiv \nu%\langle (\partial u_i/\partial x_i)^2 \rangle$. 
\langle(\partial u_i/\partial x_j + \partial u_j/\partial x_i)^2\rangle)$.
Here, summation is implied over the repeated index, and $u_i$ is the velocity component in the direction $x_i$. This simple equation cannot be solved because $\langle \epsilon \rangle$ is not known in terms of $k$ (or $u_i$).

It is traditional to think that, after some initial adjustment period, the energy decays as a power law in time $t$ as
\begin{equation}
k = K t^{-n},    
\end{equation}
where $n$ is a universal constant and $K$ has the dimensions of energy 
depending on the energy initially injected, its spectral distribution, etc., and $t$ is suitably normalized. The focus of attention is the decay exponent $n$. This power-law behavior is assumed to be valid at high enough Reynolds numbers; so we exclude from consideration very large $t$ for which the energy has plummeted to such low levels that the Reynolds number becomes very small. A similar relation for the length scale $L$ of turbulence is
\begin{equation}
L = L_0t^{m},
\end{equation}
where $L_0$ is some system-specific length and the exponent $m$ is thought to be universal. Typically, $L$ is taken as the integral scale defined as $\left(3 \pi/ 4 \right) \left(\int_{0}^{\infty} \kappa^{-1}E(\kappa) d \kappa \right) / \left( \int_{0}^{\infty}
E(\kappa) d\kappa \right)$, where $E(\kappa)$ is the energy spectrum, and is the same as the integral of the longitudinal correlation function in real space. One expects, because of nonlinear interactions among the large degrees of freedom of the flow, that HIT
at high Reynolds numbers evolves into an asymptotic universal state that is independent of initial conditions, i.e., $n$ and $m$ assume universal values. 

\begin{figure}
%\begin{center}
    \includegraphics[clip,trim=00 0 0 0,width=0.5\textwidth]{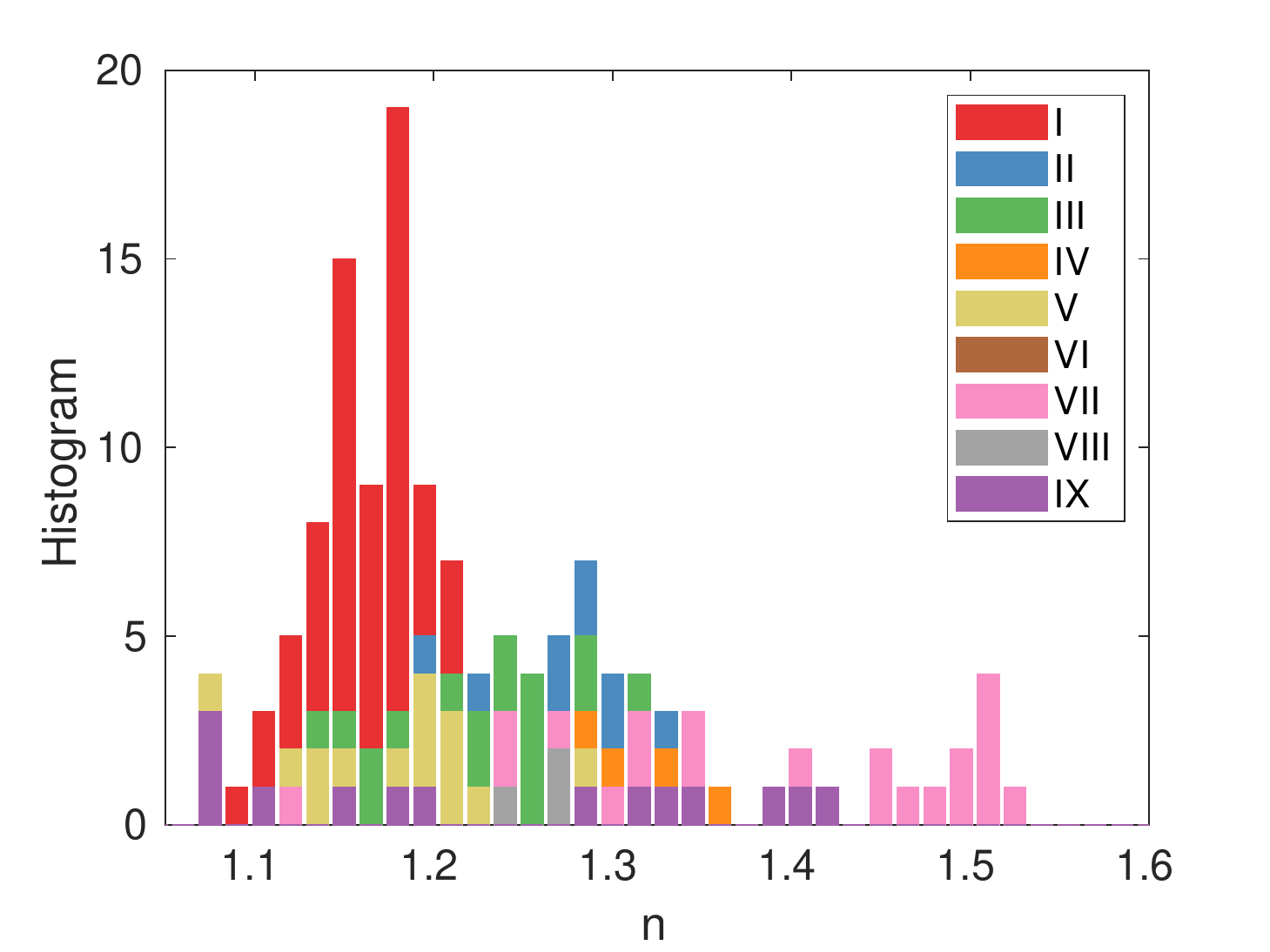}
  \includegraphics[clip,trim=00 0 0 0,width=0.5\textwidth]{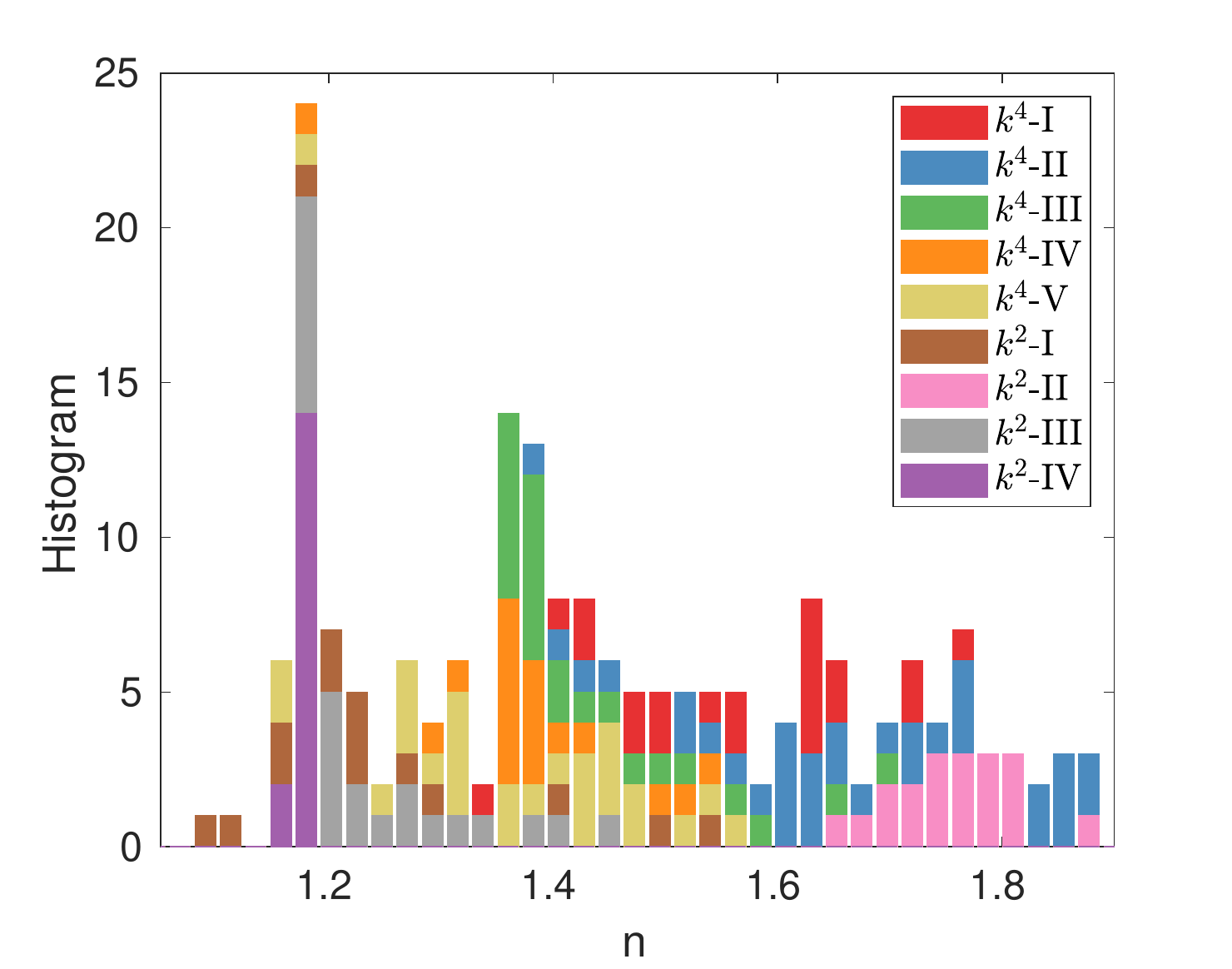}
%\end{center}
 \begin{picture}(0,0)
     \put(0,155){\small {(a)} }
\put(190,155){\small {(b)} }
\end{picture}
\caption{\label{fig:1} 
Histogram of decay exponents from the literature: (a) experiments, (b) simulations. Data in (a) are from 
I \cite{decay1}, 
II \cite{decay2a,decay2b,decay2c,decay2d}, 
III \cite{CBC66}, 
IV \cite{decay4b,decay4c,decay4d}, 
V \cite{decay5a,decay5b,decay5c,decay5d,decay5e}, 
VI \cite{decay6}, 
VII \cite{decay7}, 
VIII \cite{LW1955}, 
IX \cite{decay9a,decay9b,decay9c}. 
In (b), the initial spectrum with $E\left(\kappa\right) \propto \kappa^{4}$ near the origin correspond to: $\kappa^{4}$-I \cite{ishi2006decay,YMcPRF2018,sim1c}, $\kappa^{4}$-II \cite{set2}, $\kappa^{4}$-III \cite{set3}, $\kappa^{4}$-IV \cite{set4a,set4b,set4c,SPK2001}, $\kappa^{4}$-V \cite{set5}. The initial spectrum with $E\left(\kappa\right) \propto \kappa^{2}$ for small $\kappa$ corresponds to cases: $\kappa^{2}$-I \cite{anas2020,set6b}, $\kappa^{2}$-II \cite{sim1c}, $\kappa^{2}$-III \cite{set3}, $\kappa^{2}$-IV  \cite{set4b,set4c,set9c}. For the case \cite{SPK2001}, the simulations were compressible. Simulations here include DNS, LES using different numerical methods, and EDQNM. Some experiments and simulations are no doubt more thorough than others, but we cannot {\it a priori} discard any of them on the basis of available information. It should not be inferred that the ``correct" exponent is necessarily the most frequently observed one. }
\end{figure}
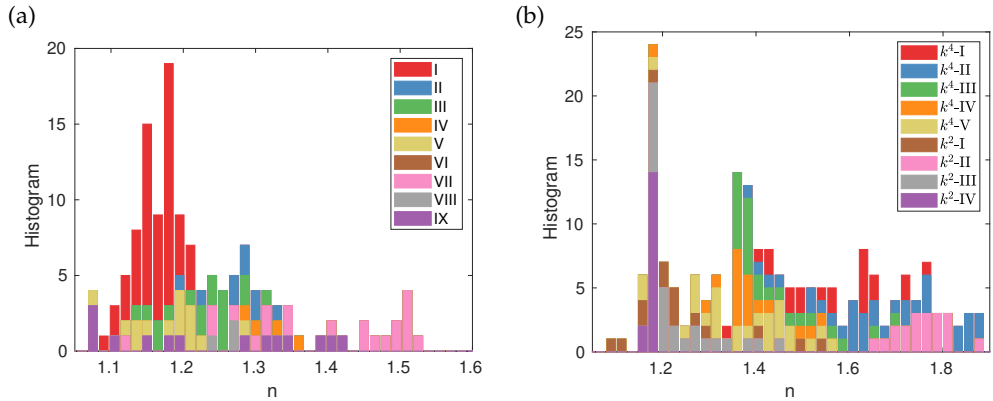

Despite the simplicity of decaying turbulence, the universality of the decay exponents has not been confirmed convincingly. We present in figure \ref{fig:1} two histograms of the decay exponent $n$ from the literature, which show a wide spread uncharacteristic of universal quantities. (A similar collection for $m$ is less useful because fewer studies have measured the length scale $L$.) Figure (a) includes data from grid experiments in wind tunnels, for which the flow far enough away from the grid is thought to be nearly homogeneous and isotropic, as well as pull-through grids in liquid tanks for which HIT is similarly attained after a certain initial adjustment time. Figure \ref{fig:1} (b) shows data from simulations. Together the two figures demonstrate that the state of decay depends, presumably, on many factors such as the Reynolds numbers, initial and residual anisotropies, boundary effects, measurements being made too close to the grid, the quality of the grid itself, the ubiquitous use of virtual origin in plotting data, inadequate ranges over which power laws are fitted, unsatisfactory simulation parameters, etc. It is clear that the problem demands closer attention. 

After a brief discussion of the basic theoretical framework for decaying turbulence, we discuss numerical simulations in section 3, present results in sections 4 and 5, and close with a brief discussion and summary of results in section 6.
 
\section{Theoretical framework}
One natural step in closing the equation at high Reynolds numbers is to represent $\langle \epsilon \rangle$ in terms of the energy introduced into the turbulent system by invoking the so-called zero-th law of turbulence: the energy injection rate at large scales equals the dissipation rate at small scales. We might take $u'$ and the integral scale $L$ as the velocity and length scales characteristic of energy injection. Writing  \begin{equation}
\langle \epsilon \rangle = C_{\epsilon} u'^{3}/L, \label{eq:diss}
\end{equation}
where $C_\epsilon$ is a constant of the order unity\footnote{$C_{\epsilon}$ depends on the Reynolds numbers when the Reynolds number is small. This dependence is reasonably well understood; see Refs. \cite{sreeni1984,DF2002, DSY2005,vandoorn}. Even in the high Reynolds number limit $C_{\epsilon}$ depends modestly on initial conditions \cite{sreeni1984}.}, (1.1) becomes
\begin{equation}
 \frac{3}{2}\frac{du'^2}{dt} = - C_{\epsilon} u'^{3}/L,
 \label{eq:kdecay1}
\end{equation}
and can be solved readily if we specify the growth of the length scale. As one point of reference, for a confined apparatus in which the length scale saturates at the size of the apparatus containing the turbulence (this could be the size of the computational box in DNS studies), we can solve \req{kdecay1} to obtain
\begin{equation}
    k \sim t^{-2} \hspace{2cm}  L \sim t^{0}.
\end{equation}
There are experiments that confirm this result \cite{vandoorn,mrsmith,srstalp}. In particular, the towed-grid experiments in a water tank \cite{vandoorn} showed the early part of the decay (50 < tU/M < 1,000) yielded an $n$ of 1.1 whereas, as the tank walls started to inhibit the growth of the integral scale, $n$ grew to 1.5 for 1,000 < tU/M < 10,000); here $t$ is measured from the time that the grid of mesh size $M$ is pulled with a velocity $U$ through the observational point.   

As another point of reference, the appealing idea of self-similarity \cite{batch1953} yields 
\begin{equation}
k \sim t^{-1} \hspace{2cm} L \sim t^{1/2}. \end{equation}
Relations (2.3) and (2.4) happen to roughly bound the extreme values observed in experiments and simulations. 

Other power laws can be deduced on the assumption that certain invariants of the K\'arm\'an-Howarth equations exist. The most important invariants are related to the spectral behavior near the wavenumber origin (or the behavior of correlation functions for large $r$). We may write the three-dimensional energy spectrum at low wavenumbers in the form \cite{batch1953}
\begin{subequations}
\begin{equation}
E(\kappa,t) = \frac{2}{\pi} I_{S} (t) \kappa^{2} + \frac{1}{3\pi} I_{L} (t) \kappa^{4} + ....., 
\label{eq:spec}
\end{equation}
where $I_{S}$ and $I_{L}$, known respectively as the Birkhoff-Saffman \cite{birkhoff1954,saffman1967large} and Loitsiansky \cite{MY.II} invariants, are defined as
\begin{equation}
%I_{BS} =  \int_{0}^{\infty} r^{2} f(r,t) dr,
I_{BS} = u'^{2} \int_{0}^{\infty} r^{2} \frac{1}{2r^{2}} \frac{\partial}{\partial r} \left[ r^{3} f(r,t)  \right] dr 
\label{eq:saff}
\end{equation}
and 
\begin{equation}
I_{L} =  u'^{2} \int_{0}^{\infty} r^{4} f(r,t) dr.
\label{eq:loit}
\end{equation}
\end{subequations}
In \req{saff} and \req{loit}, $f$ is the longitudinal correlation function that depends solely on the scalar value of the separation distance, $r$. Two classical theories for decaying turbulence based on the existence of one or the invariants have been proposed. If $I_{BS} = 0$ \cite{batch1953}, we have $E\left( \mathbf{\kappa} \right) \propto \kappa^{4}$ near the origin; in this case, $I_{L}$ is the invariant of motion, interpreted as the conservation of angular momentum \cite{david2015turb}. However, for certain initial conditions where the correlation does not decay faster than $r^{-3}$ at large scales, 
Birkhoff \cite{birkhoff1954} and Saffman \cite{saffman1967large} independently argued that $I_{L}$ diverges and $I_{BS}$ is the invariant of motion. This corresponds to the case $E\left( \kappa \right) \propto \kappa^{2}$ at the origin, and is interpreted as a result of conservation of linear momentum \cite{david2015turb}.
 
These constraints on the dynamical motion of turbulence can be exploited to obtain the power law relation for $u'^{2}$ and $L$. 
%The existence of the Birkoff-Saffman invariant corresponds to $n = 6/5$, $m = 2/5$ (see Refs.\ \cite{birkhoff1954,saffman1967large}), while that of the Loitsiansky invariant yields $n = 10/7$ and $m = 2/7$ (see Refs.\ \cite{KOL1941DEG,CBC66}).
The existence of these invariants,
can be shown  \cite{birkhoff1954,saffman1967large,KOL1941DEG,CBC66}) 
to lead, in each case, to:
\begin{align}
\text{Birkhoff-Saffman: } & \ \
n = 6/5,\ \  m = 2/5  \nonumber \\
\text{and ~Loitsiansky: } & \ \ 
n = 10/7,   m = 2/7.   \nonumber  
\end{align}

There is no obvious way to know if any of these constraints applies {\it a priori} to given experimental conditions. As a practical matter, the integrals in (2.5b) and (2.5c) can extend only as far the size of the apparatus. Looking at the histogram in \rfig{1}, one comes to the conclusion that the most likely decay exponent may be about 1.2 in both experiment and simulations, but no convincing effort has been made in those instances to relate that exponent to the existence of the Birkhoff-Saffman invariant. Thus, it is unclear at this point if there is a universal power law exponent (indeed, whether there exists a power law), and if the theory based on invariants plays a role in the experiment.  

On the other hand, in numerical simulations, one can impose either $E(\kappa) \propto \kappa^{2} \textrm{ or } \kappa^{4}$ at low wavenumbers (limited, of course, by the inverse box size) as the initial condition for the energy spectrum. In such cases, the state of turbulence may well be constrained by the invariants due to the specified spectral shape at low wavenumbers, and one may explore whether the decay exponent $n$ turns out to be $10/7$ or $6/5$, depending on whether $E(\kappa) \propto \kappa^{4}$ or  $E(\kappa) \propto \kappa^{2}$ near the origin. This is a specific thrust of the present work. 

\begin{figure}
\begin{center}
   
       \includegraphics[clip,trim=00 0 0 0,width=0.75\textwidth]{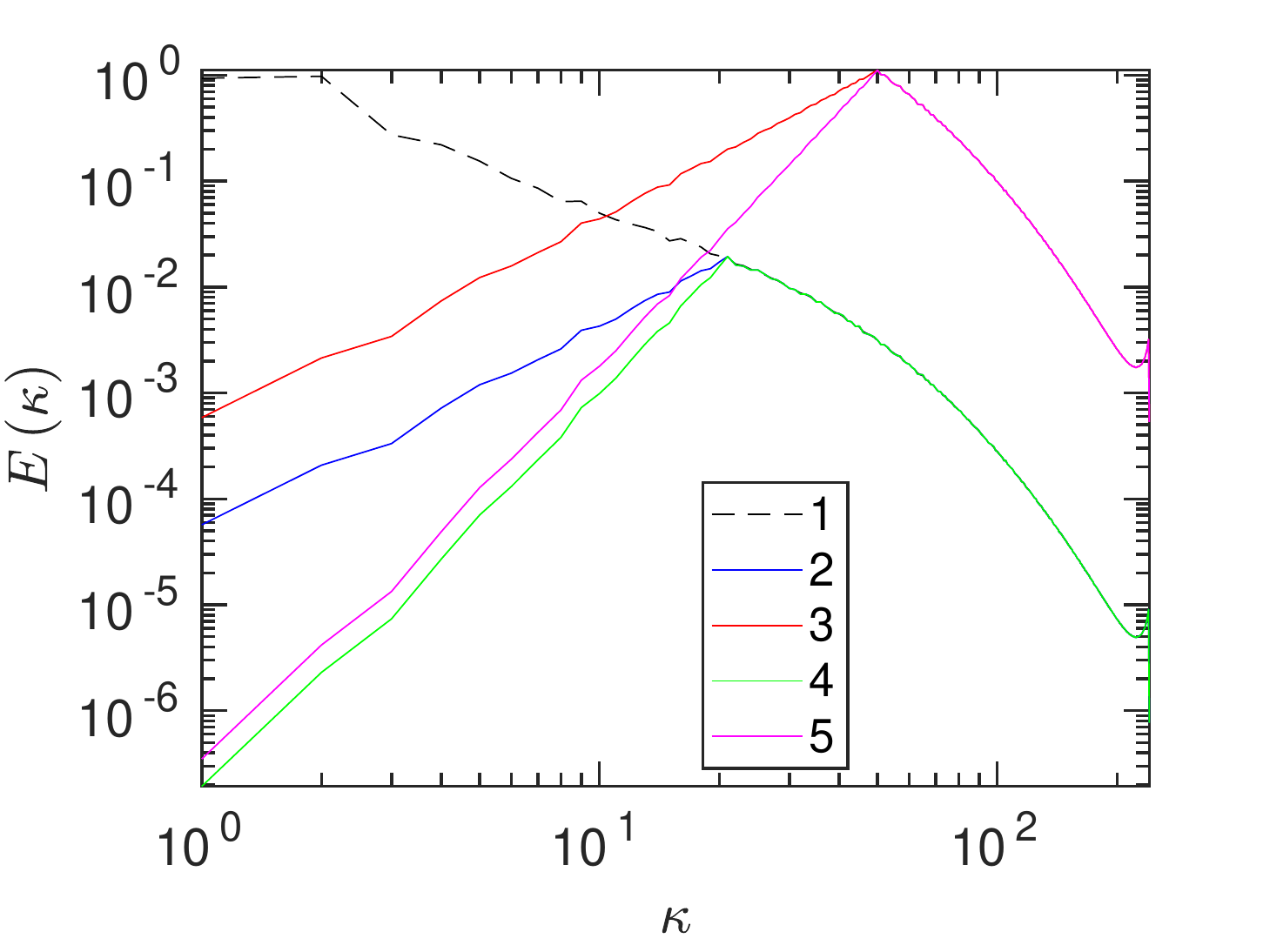}
 \end{center}
\caption{\label{fig:2} Initial spectra. 
(1) Dashed black line: energy spectra from  fully developed forced turbulence simulation; 
(2) Blue line: modified initial energy spectra with $\kappa^{2}$ scaling and $\kappa_{p} = 21$; 
(3) Red line: modified initial energy spectra with $\kappa^{2}$ scaling and $\kappa_{p} = 50$, translated upwards to increase initial Reynolds number.
(4)  Green line: modified initial energy spectra with $\kappa^{4}$ scaling and $\kappa_{p} = 21$; 
(5)  Magenta line: modified initial energy spectra with $\kappa^{4}$ scaling and $\kappa_{p} = 50$ translated upwards to increase the initial Reynolds number. See text for explanations.
  }
\end{figure} 

\section{Numerical simulations and turbulence parameters}
We conduct direct numerical simulations of turbulence in periodic boxes for a wide variety of initial conditions. The code uses a pseudospectral formulation in space and a second-order Runge-Kutta scheme for time integration. In all simulations, we generate random Gaussian-distributed velocity field as initial conditions. In DNS studies of decaying turbulence \cite{anas2020,ishi2006decay,YMcPRF2018}, it is customary to use an initial energy spectrum of the form, 
$E(\kappa) \propto \left( \kappa/\kappa_{p} \right)^{q} \exp\left[ -\left( \kappa/\kappa_{p} \right)^{2}  \right] $ 
with the power law of $q = 2 \textrm{ or } 4$ dominating low wave numbers, $\kappa < \kappa_{p}$. In this work, we modify the energy spectra of forced fully developed turbulence at low wave numbers such that, for $\kappa < \kappa_{p} $, the initial energy spectrum has the form $E \left( \kappa \right) \propto \kappa^{q}$, where $q = 2 \textrm { or } 4$. For $\kappa > \kappa_{p} $, the initial energy spectrum has the same form as that of fully developed turbulence in forced simulations. The spectrum is adjusted to be continuous at $\kappa=\kappa_{p}$. This modification results in lower initial Reynolds number compared to the forced stationary case. To increase the initial Reynolds number in some cases, we translated the modified entire energy spectrum upwards such that the initial Taylor Reynolds number is nominally comparable to that in forced stationary case in the same box. (For reference, the Taylor microscale Reynolds number $R_\lambda = u'\lambda/\nu$ where $\lambda = \sqrt{u^{'2}/\langle (\partial u/\partial x)^2} \rangle$, and $\nu$ is the kinematic viscosity.) A few initial energy spectra along with an example of the stationary case are shown in figure \ref{fig:2}. 

The relative value of $k_{p}$ is an important factor in determining the behaviour of energy decay; in our study, it has the range $5-80$ in units of the smallest allowed wave number $\kappa_{min} = 2 \pi/L_{box}$, which is unity for all simulations. For all cases, the integral length scale remains smaller than about 10\% of the box size. Simulations were made on grid sizes ranging from $256^{3} \textrm{ to } 1024^{3}$, resulting in $R_\lambda$ ranging from 23 to 456. The simulation conditions are summarized in table \ref{tab:cas4}.

\begin{table}
  \begin{center}
\begin{tabular}{|c|ccc |cccc|cccc|}
 \multicolumn{3} {c} {} & 
    \multicolumn{4} {c} {\hfill $\mathbf{\kappa^{4} \textrm{ cases }  }$   \hfill} &  
    \multicolumn{4} {c} {} \\
    
      \noalign{\vskip 3mm}    

\hline
 Cases & $R_{\lambda,0}$ & $R_{L,0}$ & $R_{L,0}/R_{\lambda,0}^{2}$ &   $\kappa_{p}$  & $\kappa_{b}$ & $\kappa_{p}/\kappa_{b}$ & $\kappa_{max}$ & $\nu$ & $L_{0}$ & $N$ & \\ 
 \hline

 \hline
  1.1 & 23 & 33 & 0.062 & 10 & 11.71 & 0.85 & 127 & 0.0031 & 0.154 & 256& \\ 
  1.2 & 60 & 78 &  0.022 &21 &  11.71 & 1.79 & 127 & 0.0031 & 0.099 & 256& \\  
 1.3 & 65 & 168 & 0.039 & 5 &  24.27 & 0.21 & 255 & 0.0011 & 0.221 & 512& \\  
1.4 & 43 & 83 & 0.045 &10 &  24.27 & 0.41 & 255 & 0.0011 & 0.135 &  512&\\  
  1.5 & 25 & 39 & 0.062 & 21  & 24.27 & 0.87 & 255 & 0.0011 & 0.085 & 512& \\  
1.6 & 180 & 335 & 0.010 & 10  & 24.27 & 0.41 & 255 & 0.0011 & 0.127 & 512& \\ 
1.7 & 155 & 242 & 0.010 & 21 & 24.27 & 0.87 & 255 & 0.0011 & 0.084 & 512& \\ 
1.8 & 142 & 183 & 0.009 & 50  & 24.27 & 2.06 & 255 & 0.0011 & 0.043 & 512& \\ 
1.9 & 77 & 94 & 0.016 & 80  & 24.27 & 3.29 & 255 & 0.0011 & 0.029 & 512 & \\  
1.10 & 9 & 7 & 0.168 & 50  & 24.27 & 2.06 & 511 & 0.0011 & 0.043 & 512 & \\ 
1.11 & 21 & 17 & 0.075 & 50  & 24.27 & 2.06 & 511 & 0.0011 & 0.043 & 512 & \\
1.12 & 45 & 87 & 0.043 & 21  & 48.47 & 0.43 & 511 & 0.0004 & 0.069 & 1024 & \\  
1.13 & 70 & 184 & 0.038 & 10  & 48.47 & 0.21 & 511 & 0.0004 & 0.116 & 1024 & \\  
1.14 & 422 & 1171 & 0.010 & 10  & 48.47 & 0.21 & 511 & 0.0004 & 0.127 & 1024 & \\  
1.15 & 383 & 566 & 0.004 & 50 & 48.47 & 1.03 & 511 & 0.0004 & 0.037 & 1024 & \\  
1.16$^{*}$ & 88 & 111 & 0.014 & 80  & 48.47 & 1.65 & 511 & 0.0002 & 0.031 & 1024 & \\  
 \hline
 \noalign{\vskip 3mm}    
   \multicolumn{3} {c} {} & 
    \multicolumn{4} {c} {\hfill $\mathbf{\kappa^{2} \textrm{ cases }  }$   \hfill} &  
    \multicolumn{4} {c} {} \\
  \noalign{\vskip 3mm}

\hline
 Cases & $R_{\lambda,0}$ & $R_{L,0}$  & $R_{L,0}/R_{\lambda,0}^{2}$ &  $\kappa_{p}$ &  $\kappa_{b}$ & $\kappa_{p}/\kappa_{b}$& $\kappa_{max}$ & $\nu$ & $L_{0}$ & $N$& \\ 
 \hline
 2.1 & 26 & 41 & 0.061 & 10 & 11.71 & 0.85 & 127 & 0.0031 & 0.181 & 256& \\  
  2.2 & 71 & 101 & 0.020 & 21  & 11.71 & 1.79 & 127 & 0.0031 & 0.116 & 256&\\  
 2.3 & 70 & 203 & 0.041 & 5 &  24.27 & 0.21 & 255 & 0.0011 & 0.257 & 512& \\  
2.4 & 46 & 101 & 0.048 & 10 & 24.27 & 0.41 & 255 & 0.0011 & 0.157 & 512& \\  
  2.5 & 28 & 49 & 0.063 & 21 &  24.27 & 0.87 & 255 & 0.0011 & 0.099 & 512& \\  
2.6 & 197 & 424 & 0.011 &  10 &  24.27 & 0.41 & 255 & 0.0011 & 0.157 & 512& \\  
 2.7 & 175 & 304 & 0.010 &21 &  24.27 & 0.87 & 255 & 0.0011 & 0.099 & 512& \\    
  2.8 & 170 & 241 & 0.008 &  50 &  24.27 & 2.06 & 255 & 0.0011 & 0.051 & 512& \\  
  2.9 & 95 & 127 & 0.014 & 80 &  24.27 & 3.29 & 255 & 0.0011 & 0.034 & 512& \\  
  2.10 & 13 & 9 & 0.165 & 50 &  24.27 & 2.06 & 511 & 0.0011 & 0.051 & 512& \\ 
2.11 & 28 & 20 & 0.070  & 50 &  24.27 & 2.06 & 511 & 0.0011 & 0.051  & 512& \\
  2.12 & 49 & 107 & 0.044 &  21 &  48.47 & 0.43 & 511 & 0.0004 & 0.081 & 1024& \\ 
  2.13 & 76 & 231 & 0.040 & 10 &  48.47 & 0.21 & 511 & 0.0004 & 0.140 & 1024& \\ 
  2.14 & 456 & 1476 & 0.007 &  10 &  48.47 & 0.21 & 511 & 0.0004 & 0.154 & 1024& \\ 
  2.15 & 436 & 716 & 0.004 & 50 &  48.47 & 1.03 & 511 & 0.0004 & 0.043 & 1024& \\ 
  2.16$^{*}$ & 373 & 535 & 0.004 & 80 &  48.47 & 1.65 & 511 & 0.0002 & 0.046 & 1024& \\ 
  \hline 
  \end{tabular}
  \end{center}
\caption{Table of simulation parameters and initial conditions. Here $R_{\lambda,0}$ and $R_{L,0}$ are the initial Reynolds numbers based on the Taylor microscale and integral scale, respectively; $\kappa_{p}$ is the wavenumber up to which the initial spectrum has a prescribed power law of the form $E \left(\kappa \right) \propto \kappa^{q}$ near the origin, where $q= 2 \textrm{ or } 4$. The parameter, $\kappa_{b}$ is the bottleneck wavenumber based on forced turbulence (refer to the text for more details) and $\kappa_{max}$ is the maximum wavenumber supported by the grid. Also, $\nu$ and $L_{0}$ are the kinematic viscosity of the fluid and initial integral length of the flow. In all simulations, the lowest resolved wavenumber, $\kappa_{min}$ is $1$; $N$ is the linear grid size. The number of realizations $N(r)$ used to calculate the ensemble average is 5 for all cases, except for 1.14 and 2.16 for which it is 4. The starred $\left(^{*} \right)$ cases 1.16 and 2.16 have the initial spectrum of the form $E \left( \kappa \right) \propto \left(\kappa/\kappa_{p}\right)^q \exp\left[-\left( \kappa/\kappa_{p}\right)^{2} \right]$, for both q  = 2 and 4.}
  \label{tab:cas4}
  \end{table}

\section{Preliminary results on decay exponents}\label{sec:crit}
In experiments behind grids in wind tunnels, for which the power-law extends typically over a decade or so, researchers have often used by trial and error a virtual origin to improve the quality of power law fits \cite{decay2a}. Since we perform simulations for over a thousand integral scales, we calculate the decay exponent assuming the existence of the power law decay of the form $k \propto t^{-n}$ without invoking a virtual origin. The incorporation of a modest virtual origin makes no difference to our conclusions on energy decay. We obtain the scaling exponent locally by evaluating
\begin{equation}
  n = - \frac{ d \ln k}{d \ln t }.
 \label{eq:kt}
\end{equation}
A plateau implies the existence of a well-defined power law.

\begin{figure}
    \includegraphics[clip,trim=00 0 0 0,width=0.5\textwidth]{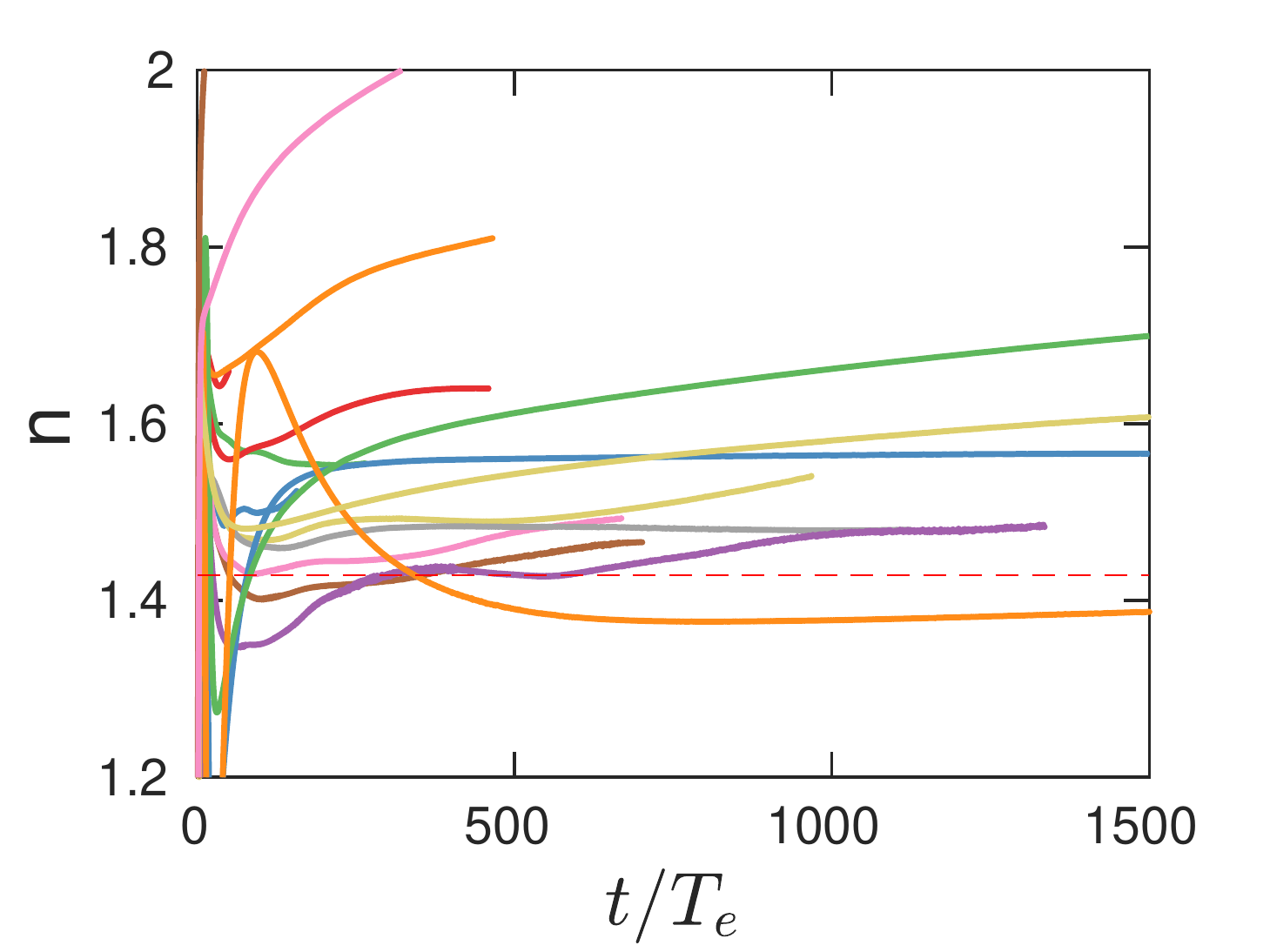}
     \includegraphics[clip,trim=00 0 0 0,width=0.5\textwidth]{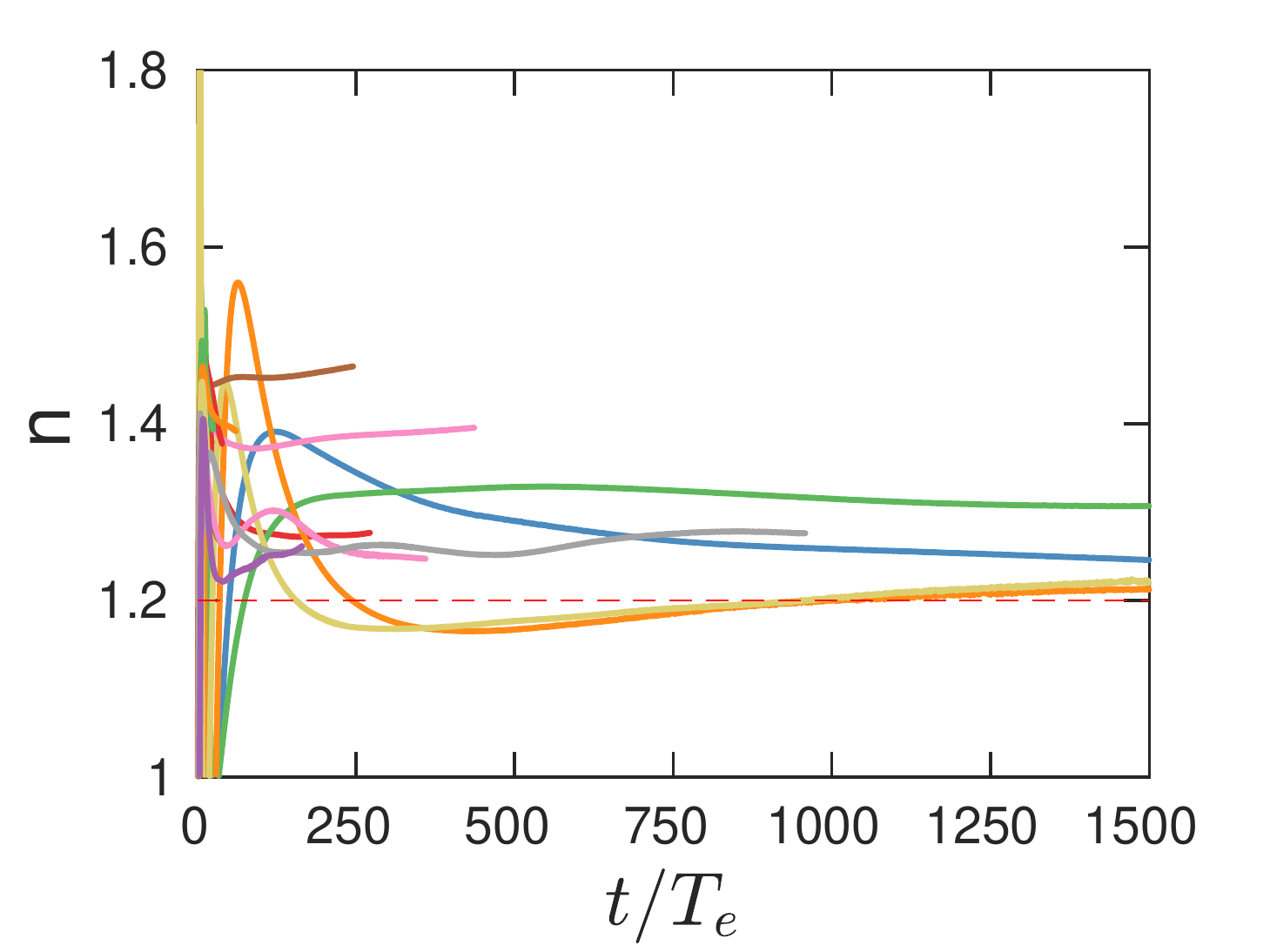}
     \begin{picture}(0,0)
\put(385,273){\small {(a)} }
\put(190,273){\small {(b)} }
%\put(-385,133){\small {(c)} }
%\put(-192,133){\small {(d)} }
\end{picture}
 \begin{picture}(0,0)
     \put(0,165){\small {(a)} }
\put(210,165){\small {(b)} }
\end{picture}
\caption{\label{fig:3} The decay exponent for all cases. Initially, (a) $E \left( \kappa \right) \propto \kappa^4 $ and (b) $E \left( \kappa \right) \propto \kappa^2 $ for small $\kappa$. The red dashed lines correspond to classical theories: 
(a)  $-10/7$, Kolmogorov's form; (b) $-6/5$, Birkhoff-Saffman form.
 }
\end{figure}
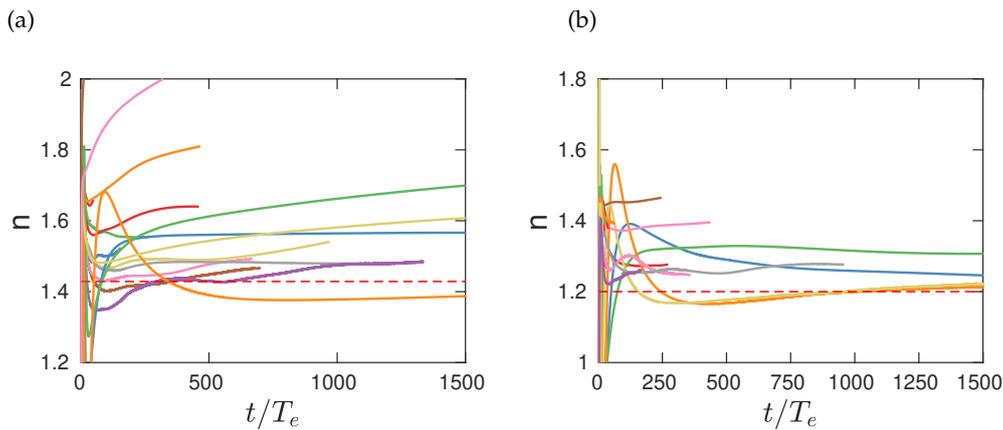 

In our first attempt, we sampled a variety of initial conditions in terms of Reynolds number and spectral form of turbulent energy, roughly mimicking the experiment, and obtained decay exponents for all cases, as displayed in figures \ref{fig:3} (a,b). Figures (a) and (b) correspond, respectively, to $\kappa^{4}$ and $\kappa^{2}$ behaviours at low wavenumbers. In both figures, time is normalized by the initial eddy turnover time, $T_{e}= L_{0}/u'_{0}$, the ratio of initial values of the integral length scale to the root-mean-square (rms) velocity. It is readily seen that not all cases have a plateau. For those with approximate plateaus, the numerical values are not all the same. Further, it seems that the plateau is sometimes achieved after a relatively long time, $t/T_{e} \ge 0(100)$ (which often corresponds to the outer edge of power-law fits in wind tunnel experiments). If one does not observe the decay for long enough times and take local slopes, it appears that the dip in early times can be construed as a power law. These slopes indeed mimic the variability in the histogram (Figs.\ 1 (a) and (b)). It must also be noted that not one of them has unity exponent (and the one case of $n$ of about $2.25$ appears to correspond to the final period of decay).

Does this mean that there is no well-defined decay exponent? It suggests, instead, that better control of conditions may be needed to be more definitive. First, we need the Reynolds numbers to be high enough during the decay process; since, for all exponents greater than unity, the Reynolds number decreases with time, this condition is especially demanding for the Kolmogorov case because of its larger decay exponent. We ought to make sure that the energy input occurs at wavenumbes sufficiently far away from the the dissipative range; otherwise, a substantial decay could occur directly at the energy-containing scale itself. Further, the length scale should not grow to be a substantial fraction of the computational box. These are often conflicting requirements, and only a few runs of table 1 satisfy these conditions, even if only approximately. We shall focus on just those cases here. Particularly relevant is the role of the postulate on the permanence of large eddies \cite{MY.II}. 

A quick examination of table 1 suggests that cases 1.6, 1.7, 1.8, 1.14 and 1.15 have the best chance of retaining high enough in Reynolds number for the $\kappa^4$ cases. Likewise, 2.6, 2.7, 2.9, 2.14 and 2.15 appear to be the most suitable among the $\kappa^2$ cases. 

The second criterion is based on the fact that the energy-containing scales must be distinct from the viscous-dominated scales of turbulence; in other words, $\kappa_{p}$ should be far enough to the left of the dissipative region. For box turbulence for which the lowest resolved wavenumber $\kappa_{min}= 1$, we stipulate as a guidance that $\kappa_{p}$ should satisfy 
\begin{equation}
\frac{\kappa_{p}}{\kappa_{b}} < 1, 
    \label{eq:criteria2}
\end{equation}
where $\kappa_b$ is the wavenumber corresponding to the bottleneck region, centered at $\kappa_{b}  = 0.13 / \eta$ \cite{donzis2012}. Here, $\eta = \nu^{3}/ \langle \epsilon_i \rangle^{1/4}$ 
is the Kolmogorov length scale from forced turbulence simulations, which satisfies the criterion that $\kappa_{max} \eta > 1.5 $ and resolves most significant scales in the problem; $\langle \epsilon_i \rangle$ is the space-time averaged dissipation in the stationary state prior to the onset of decay.

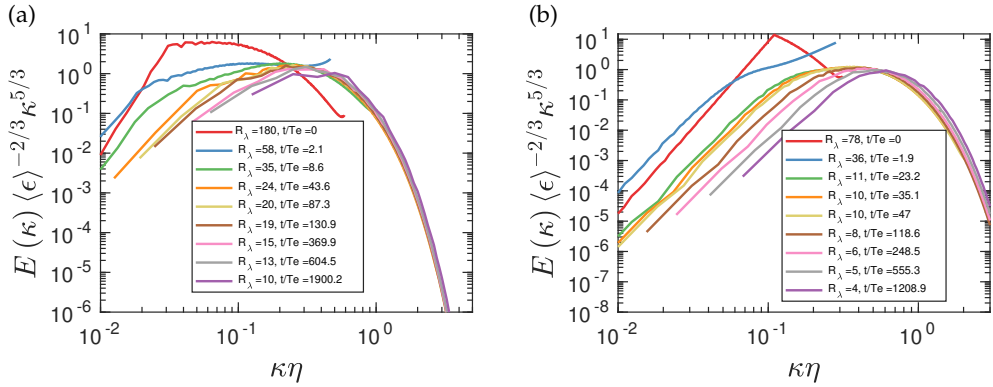
\begin{figure}
     \includegraphics[clip,trim=00 0 0 0,width=0.5\textwidth]{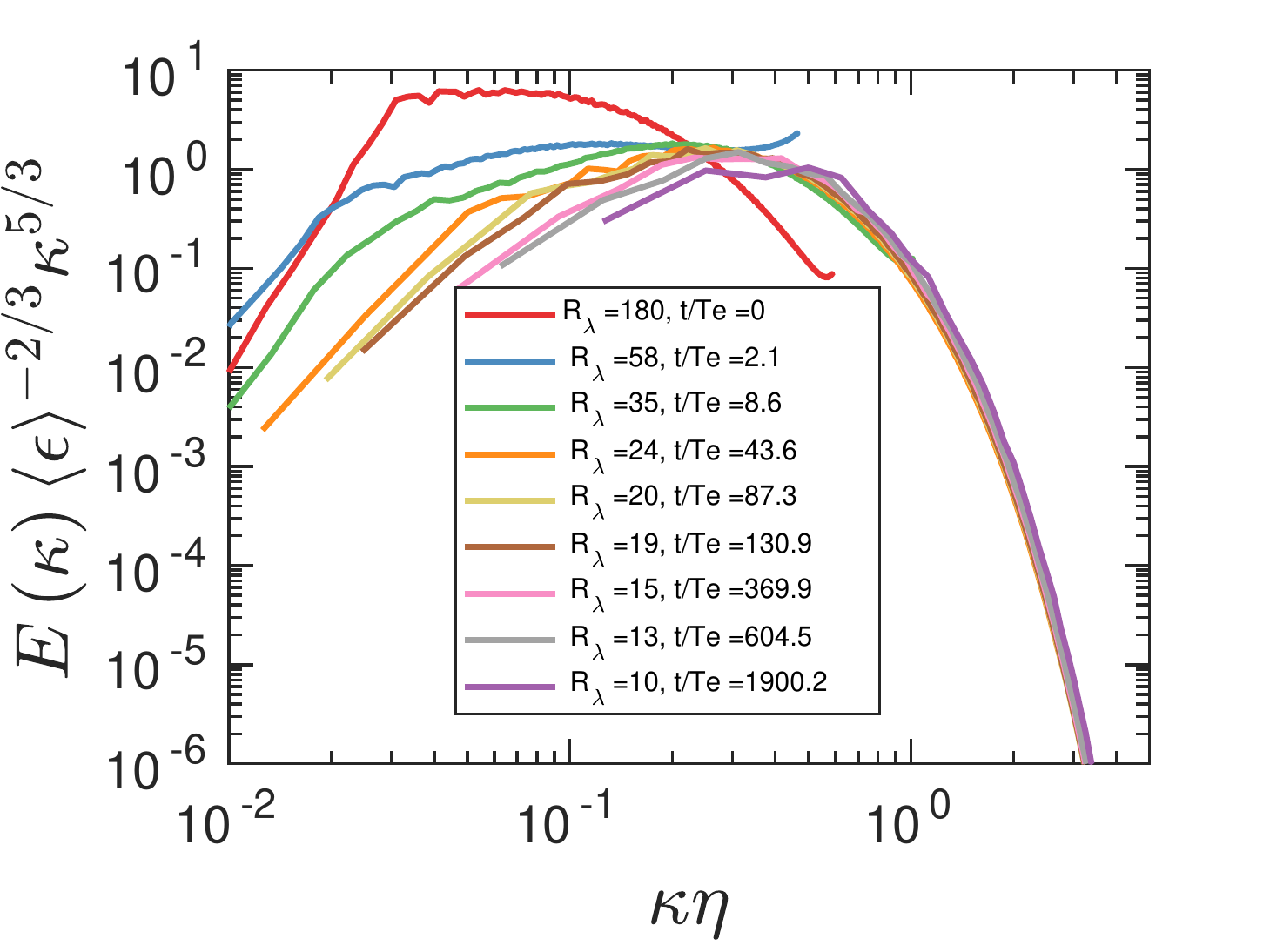}
     \includegraphics[clip,trim=00 0 0 0,width=0.5\textwidth]{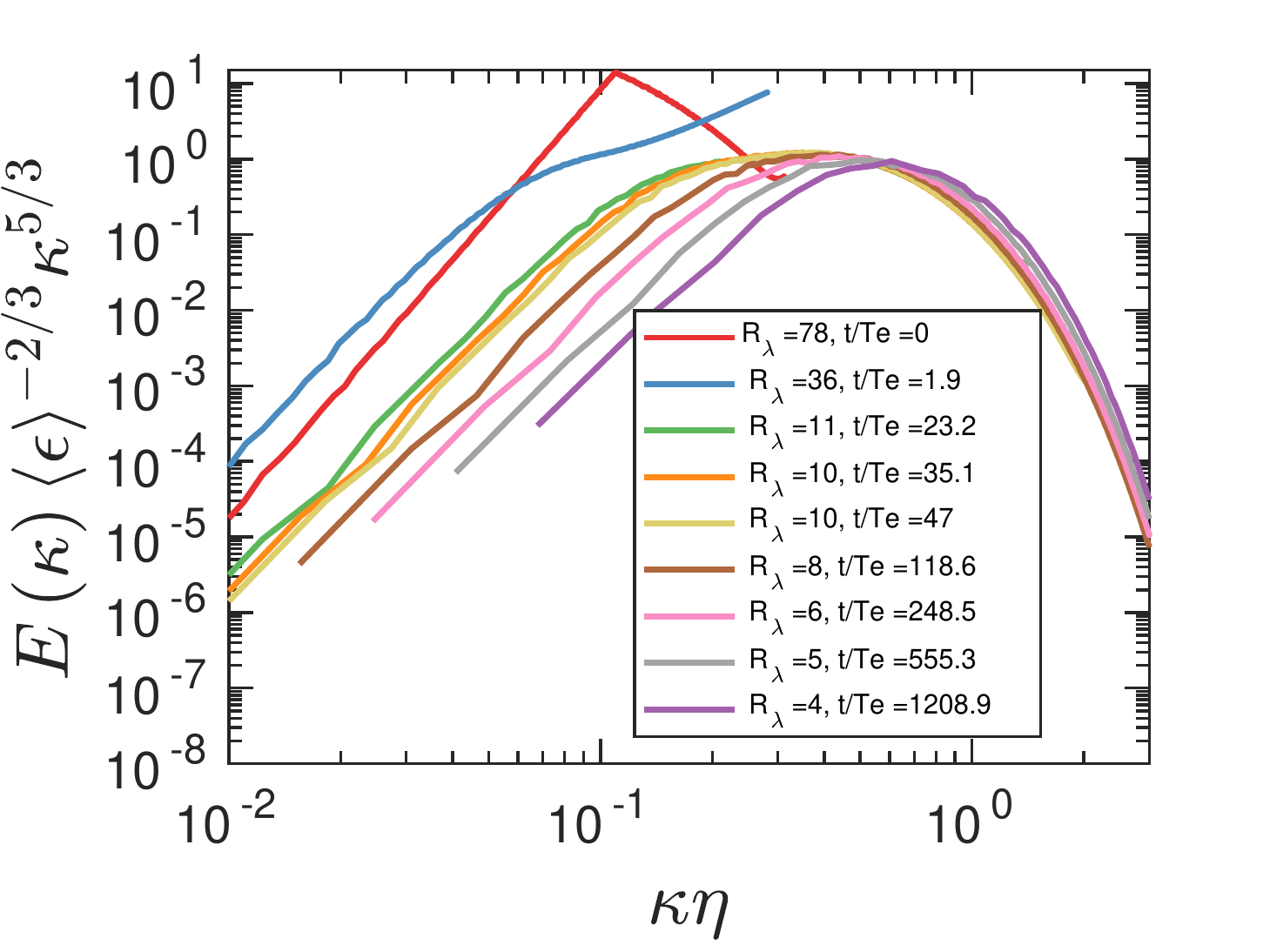}
     \begin{picture}(0,0)
       \put(0,150){\small {(a)} }
\put(195,150){\small {(b)} }
     \end{picture}
\caption{\label{fig:k4_spec} Spectra scaled on Kolmogorov variables for two typical $\kappa^{4}$ cases: (a) \textit{1.6} and (b) \textit{1.9}. (a) is one of the high Reynolds number cases but (b) is not. See table \ref{tab:cas4} and text for more details. 
 }
\end{figure}

\begin{figure}
        \includegraphics[clip,trim=00 0 0 0,width=0.5\textwidth]{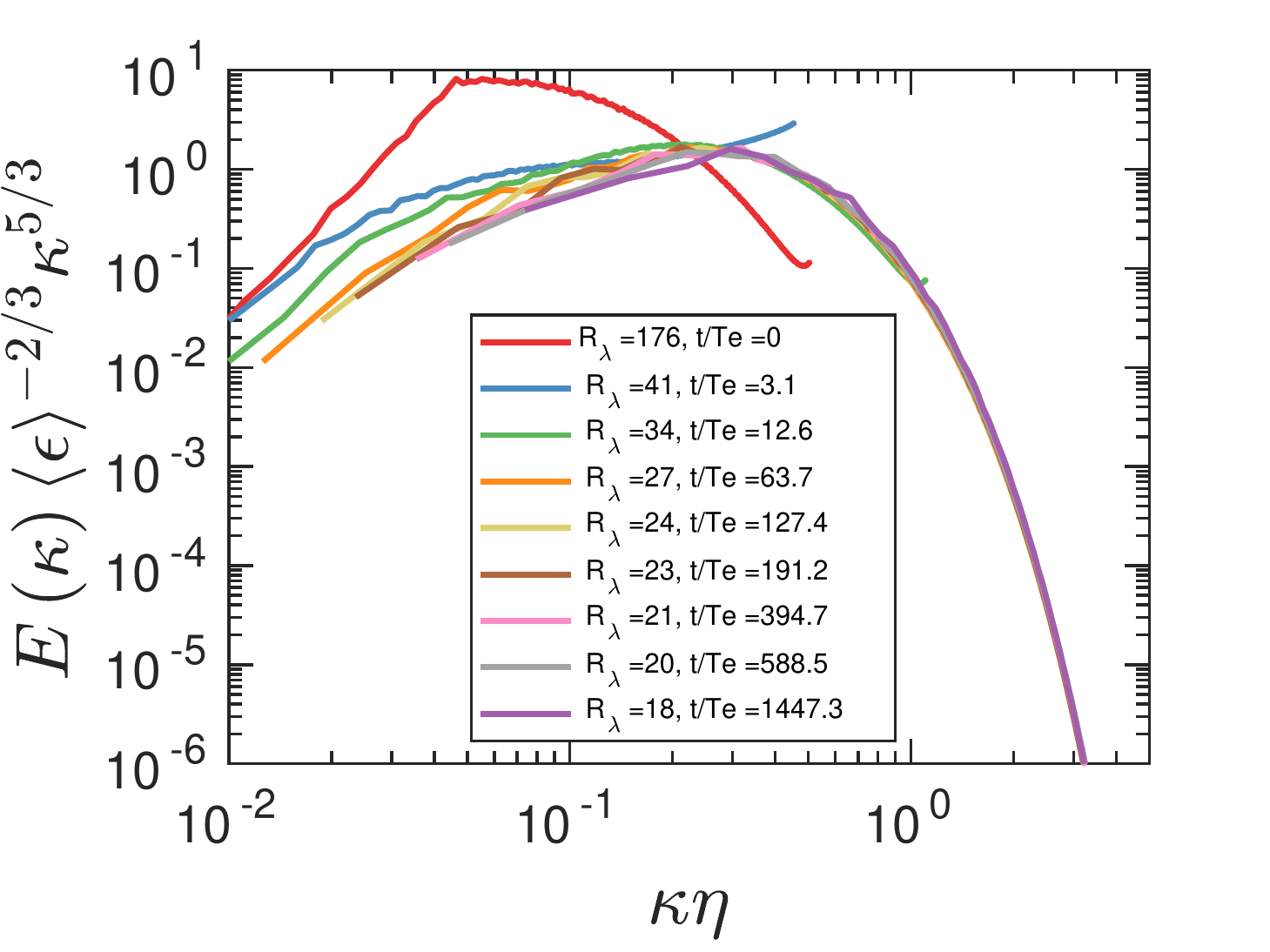}
    \includegraphics[clip,trim=00 0 0 0,width=0.5\textwidth]{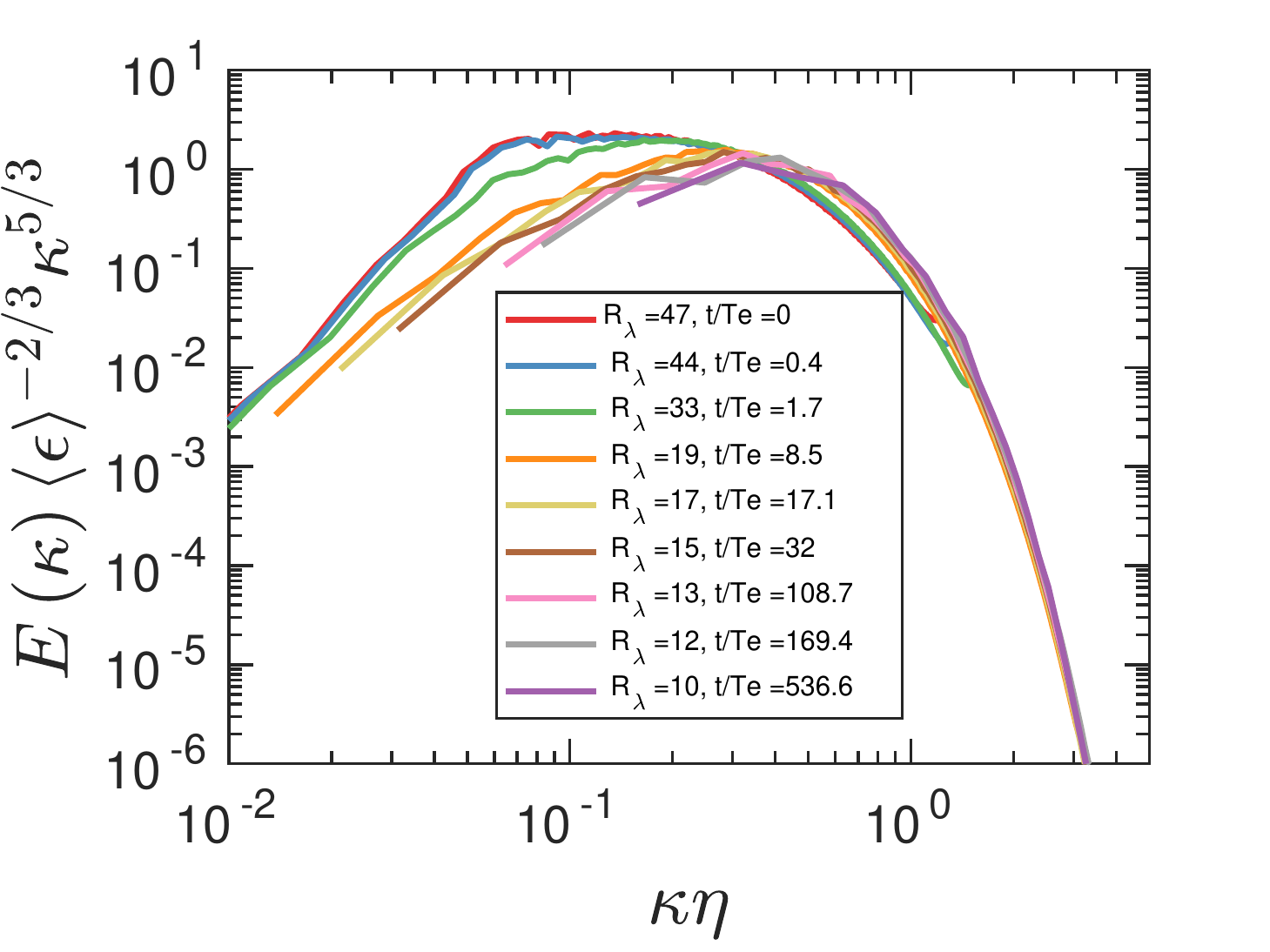}
\begin{picture}(0,0)
 \put(0,150){\small {(a)} }
\put(195,150){\small {(b)} }
   \end{picture}
\caption{\label{fig:k2_spec} Spectra scaled on Kolmogorov variables for two typical $\kappa^{4}$ cases: (a) \textit{2.7} and (b) \textit{2.4}. (a) is one of the three high Reynolds number cases but (b) is not. See table \ref{tab:cas4} and text for more details.
 }
\end{figure}

If we apply this additional criterion, it is immediately clear that 1.6, 1.7 and 1.14, as well as 2.6, 2.7 and 2.14, are the probably the best cases to consider. We will consider mostly these cases and comment on others only when appropriate. 

One general comment is useful before considering specific results. Figures \ref{fig:k4_spec} and \ref{fig:k2_spec} show the energy spectra normalized according to the Kolmogorov form for small scales. For both cases, after some initial transients, the higher-wavenumber spectra collapse to the standard accuracy (see \cite{MY.II}) under Kolmogorov's \cite{K41} similarity for $R_{\lambda} \ge 8$. A more detailed study of this feature (Refs.\ \cite{khurshid,buaria}) has shown that this universality is at best approximate, but may still be regarded as a first approximation.

\section{Results from further considerations} \label{sec:k4}
\subsection{The decay exponents}
\begin{figure}
       \includegraphics[clip,trim=00 0 0 0,width=0.5\textwidth]{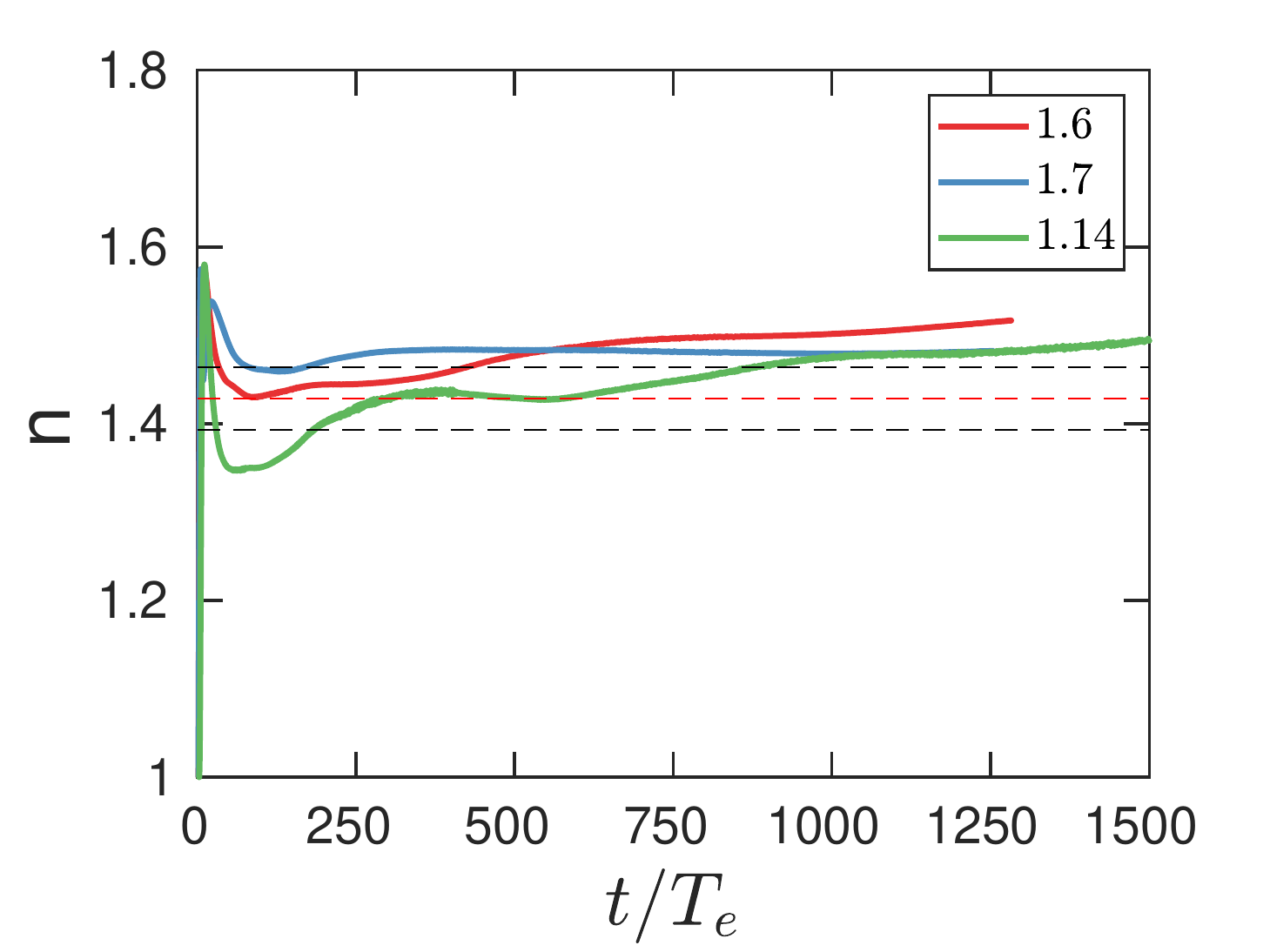}
                \includegraphics[clip,trim=00 0 0 0,width=0.5\textwidth]{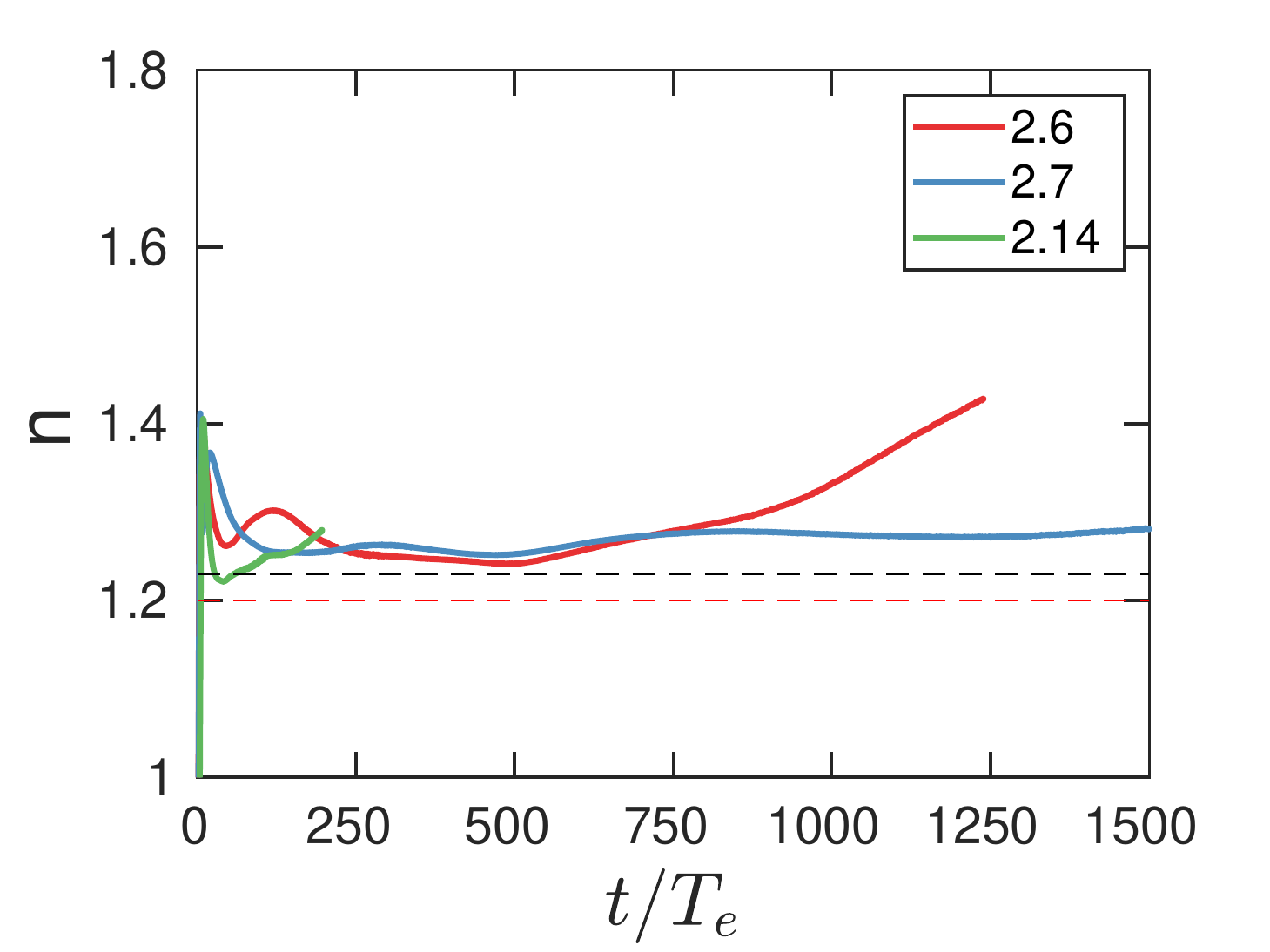}
    \begin{picture}(0,0)
 \put(0,150){\small {(a)} }
\put(195,150){\small {(b)} }
   \end{picture}
\caption{\label{fig:bestdecay} The decay exponents for all three plausible cases of (a) $k^{4}$ and (b) $k^{2}$. The red dashed lines correspond to the theoretical exponent: (a)  $n= 10/7$ and (b) $6/5$. The black dashed lines correspond to (a) $10/7 \pm 2.5\%$ and (b) $6/5 \pm 2.5\%$.
 }
\end{figure}
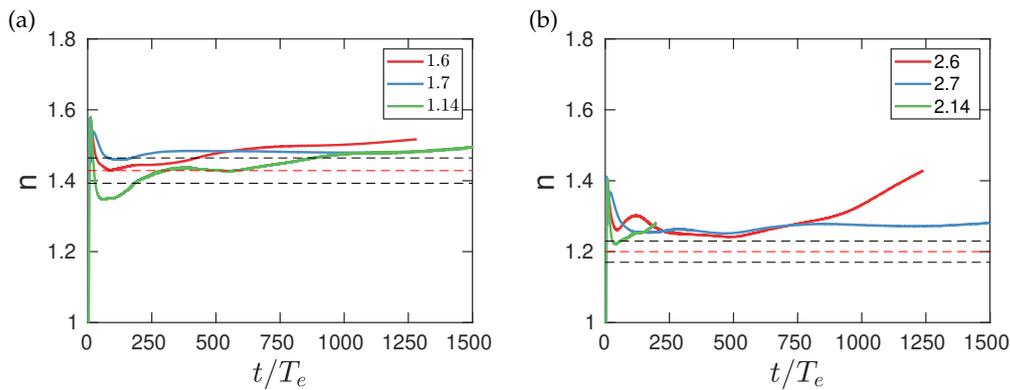

We plot in figures \ref{fig:bestdecay} the decay exponents for the chosen cases with $\kappa^4$ and $\kappa^2$ variation near the origin. The decay exponent is close the theoretical value of $10/7$ predicted by the Kolmogorov theory for the $\kappa^4$ case, and to the theoretical value of $6/5$ predicted by the Birkhoff-Saffman theory for the $\kappa^4$ case. Cases \textit{1.14} and \textit{1.7} have a convincing plateau for long periods of time although exponent for the case \textit{1.6} seems to vary slowly with time slowly for the $\kappa^4$ case; similarly for the $\kappa^2$ case, \textit{2.14} and \textit{2.7} seem to be constants even though \textit{2.6} departs from its plateau behavior somewhat abruptly for $t/T_E > 500$ or so.  
The constants 
for the $\kappa^4$ and $\kappa^2$ cases are unmistakably different.

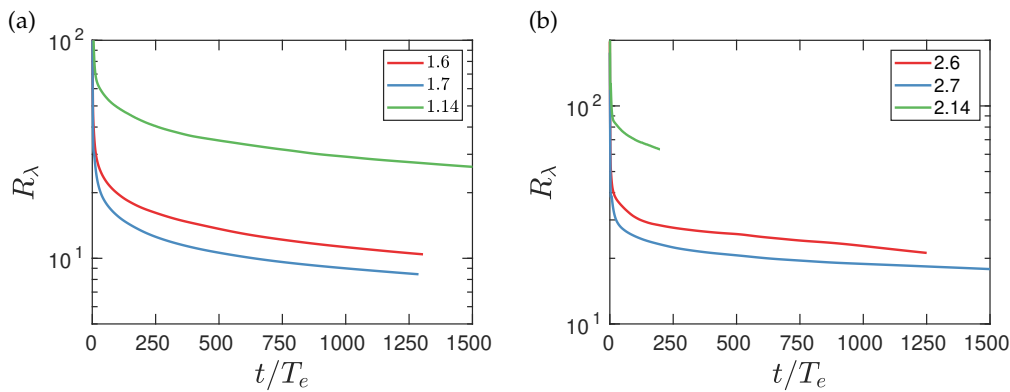
\begin{figure}
          \includegraphics[clip,trim=00 0 0 0,width=0.5\textwidth]{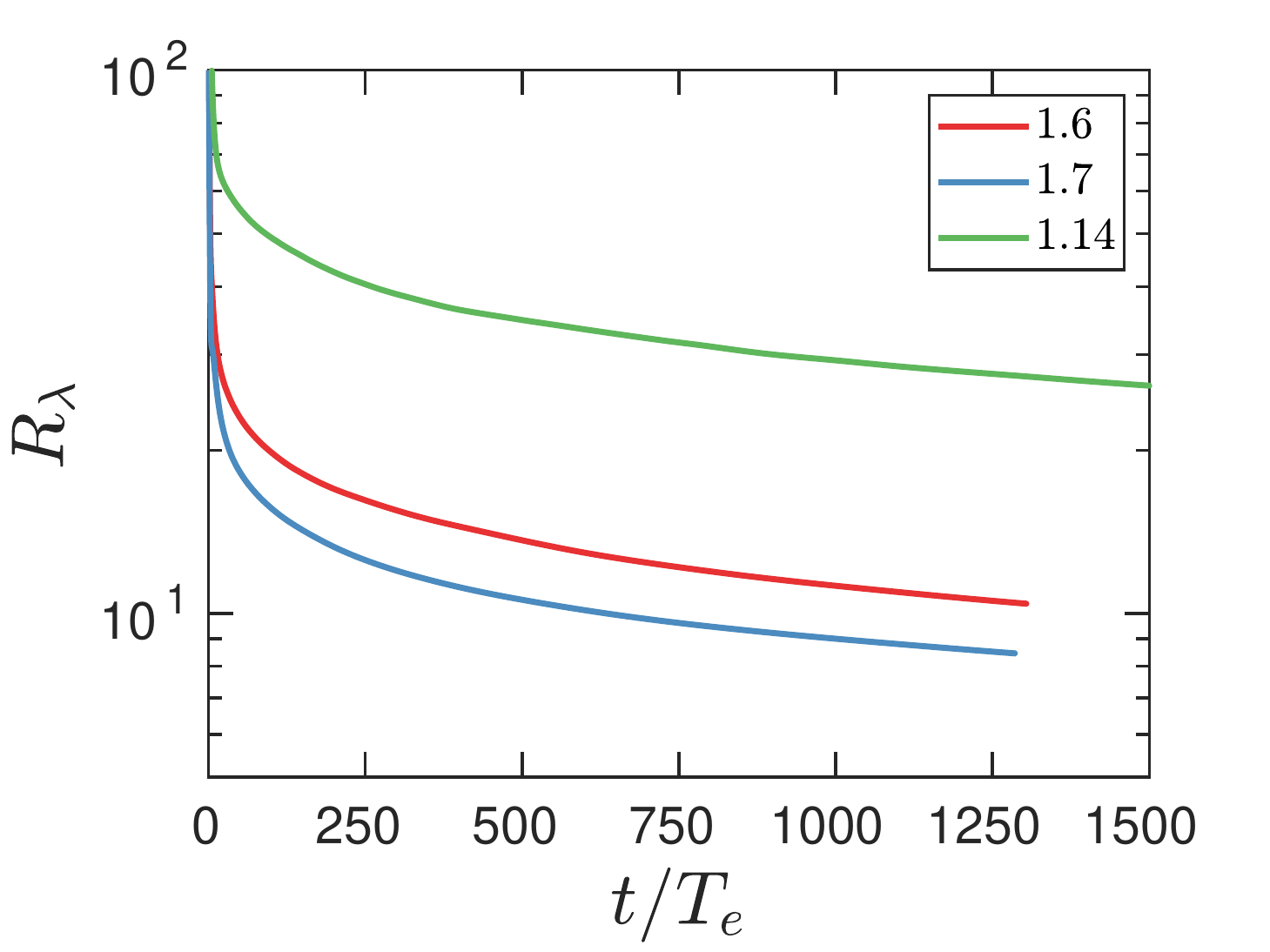}
  \includegraphics[clip,trim=00 0 0 0,width=0.5\textwidth]{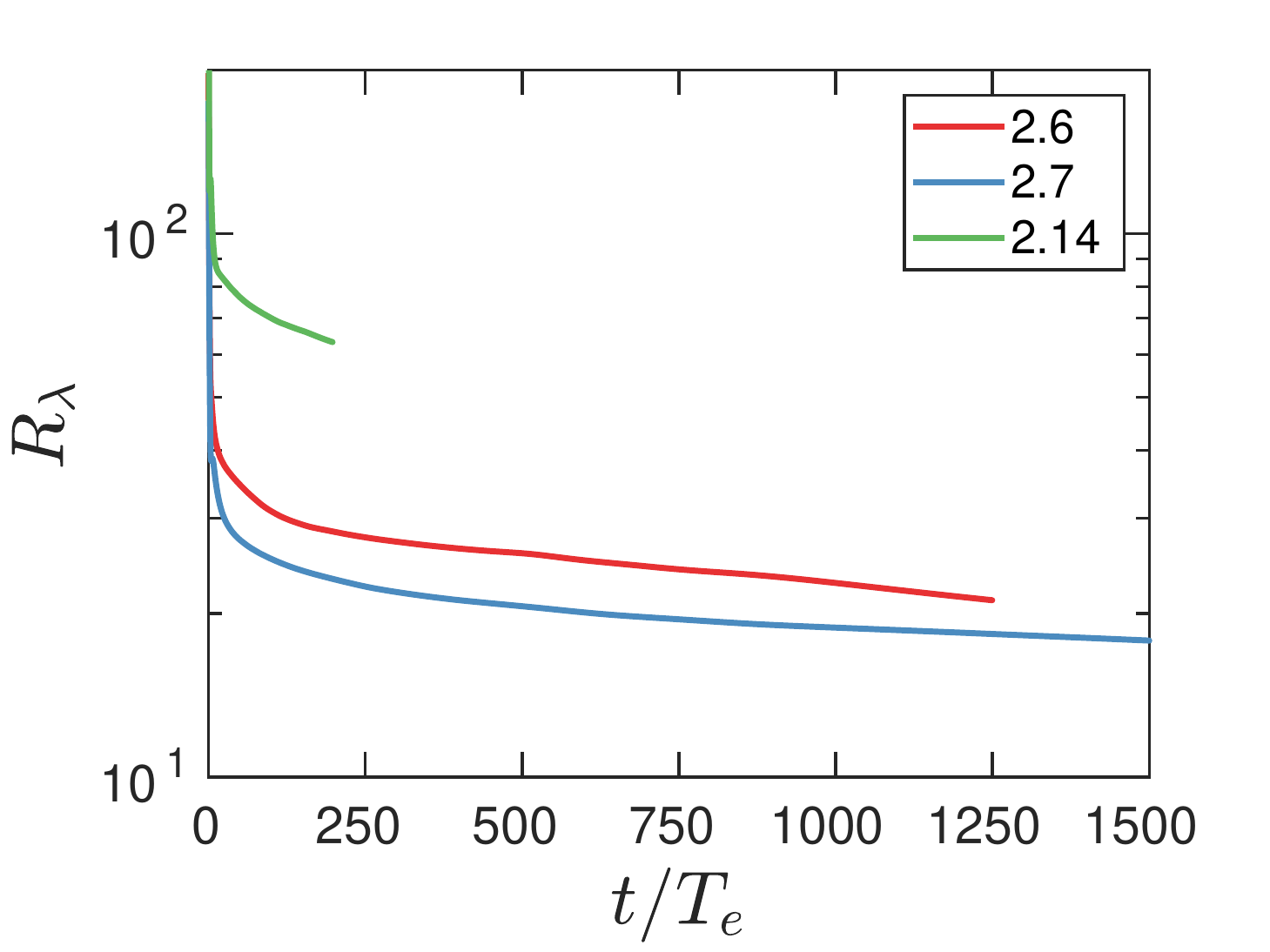}
  \begin{picture}(0,0)
 \put(0,150){\small {(a)} }
\put(195,150){\small {(b)} }
   \end{picture}
\caption{\label{fig:relam} Evolution of the Taylor microscale Reynolds number as a function of time for the best cases with (a) $k^{4} $ and (b) $k^{2}$ behaviour near the origin. 
 }
\end{figure}

\begin{figure}
          \includegraphics[clip,trim=00 0 0 0,width=0.5\textwidth]{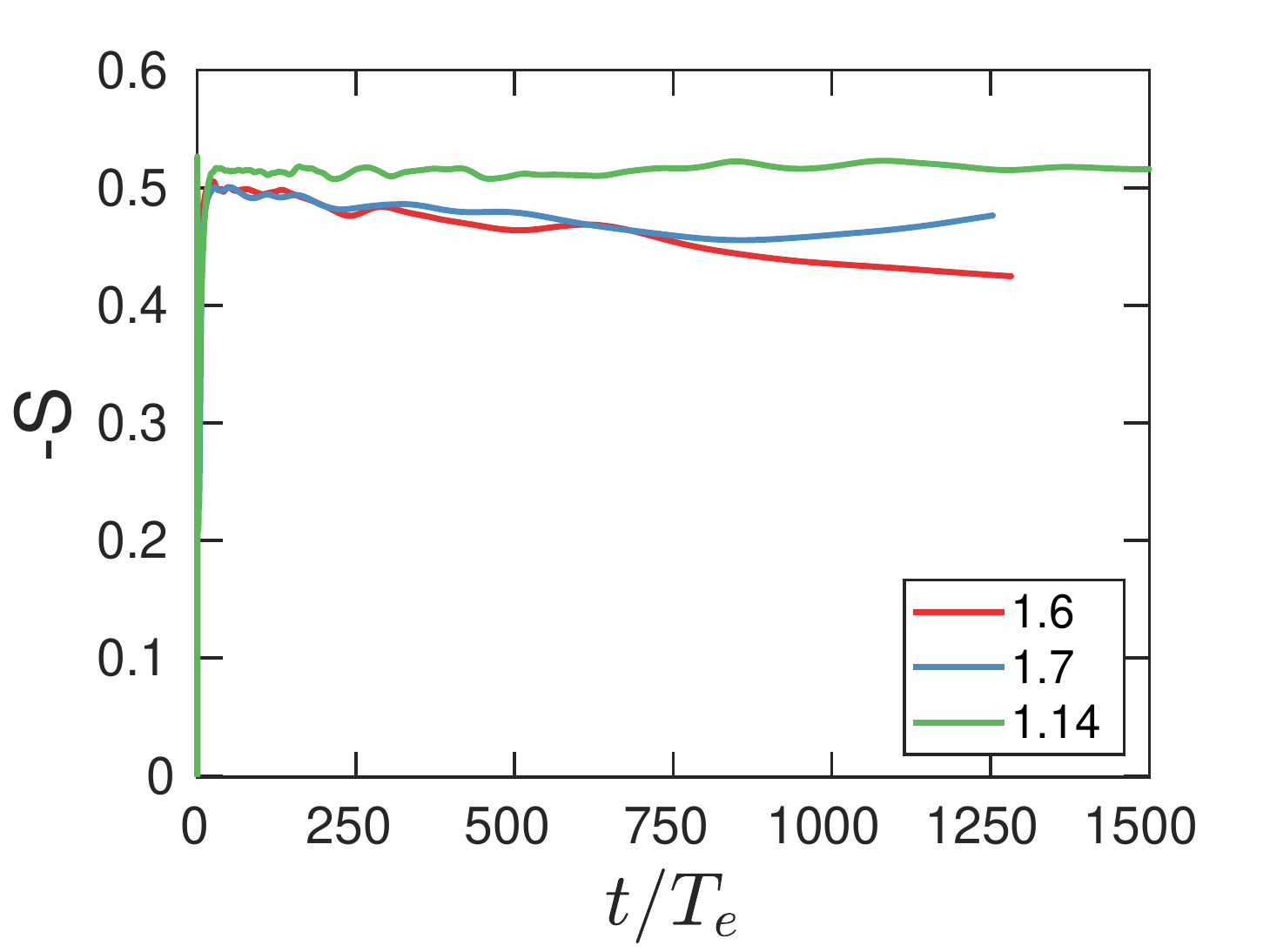}
\includegraphics[clip,trim=00 0 0 0,width=0.5\textwidth]{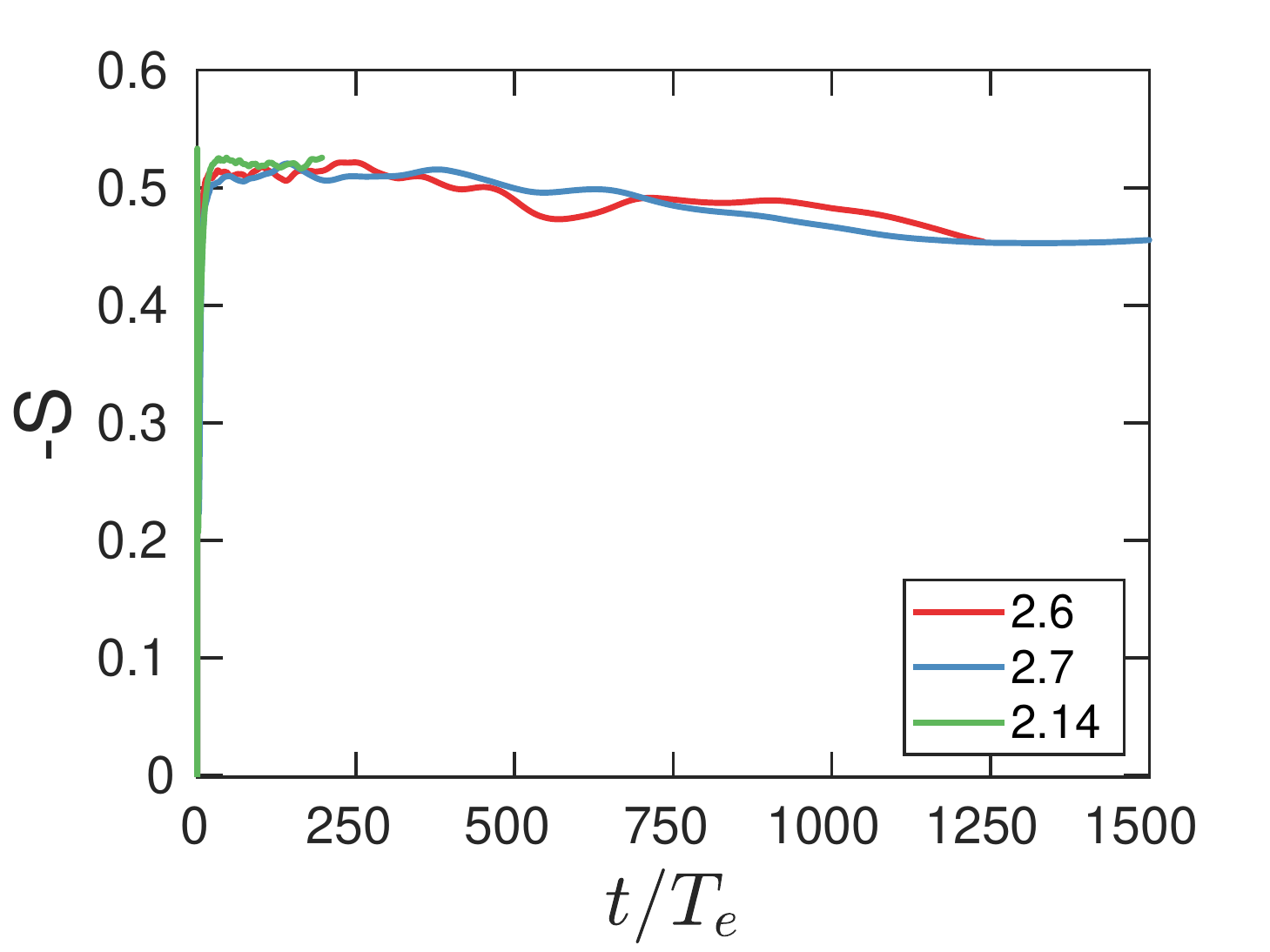}
 \begin{picture}(0,0)
  % \put(0,300){\small {(a)} }
%\put(195,300){\small {(b)} }
      \put(0,155){\small {(a)} }
\put(195,155){\small {(b)} }
 %       \put(40,190){\small {A} }
%\put(40,45){\small {B} }
%\put(235,45){\small {C} }
%\put(235,190){\small {D} }
\end{picture}
\caption{\label{fig:k4_skew} Velocity derivative skewness for the decay cases examined. The skewness is not far from the standard value of $-0.5$ expected for fully turbulent turbulence.
 }
\end{figure}
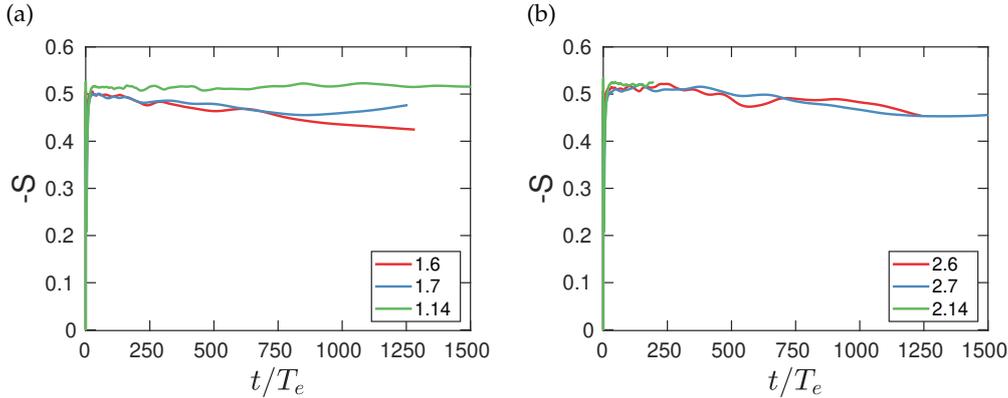

Figure \ref{fig:relam} shows that the Reynolds number remains moderate in all cases as the decay proceeds. Cases 1.7 and 2.7 have the lowest Reynolds numbers, so we may expect some deviations in other considerations below. Further, the skewness of the longitudinal velocity derivative behaves approximately as expected, at least for $t/t_E < 500$ (see Fig.\ \ref{fig:k4_skew}). An interesting feature of this figure is that, although the Reynolds numbers initially range from 155 to 422, sustained skewness of $−0.5$ is obtained only after an initial adjustment period at which point $R_\lambda$ falls to about 40 and the skewness reaches its maximum. This suggests that the decaying turbulence could have a long transient.

Taken together, the basic result appears to be that the flow fields with suitably controlled initial conditions evolve approximately into universal decay states expected by the Kolmogorov and the Birkhoff-Saffman forms. It has not been possible, despite considerable effort invested by us (see table 1 and the discussion section), to obtain closer conformity to the theories.  

\subsection{Growth of integral length scales}

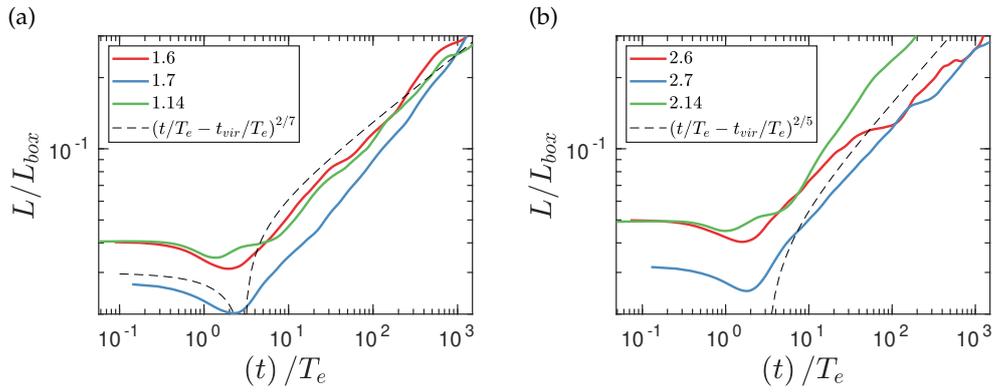
\begin{figure}
           \includegraphics[clip,trim=00 0 0 0,width=0.5\textwidth]{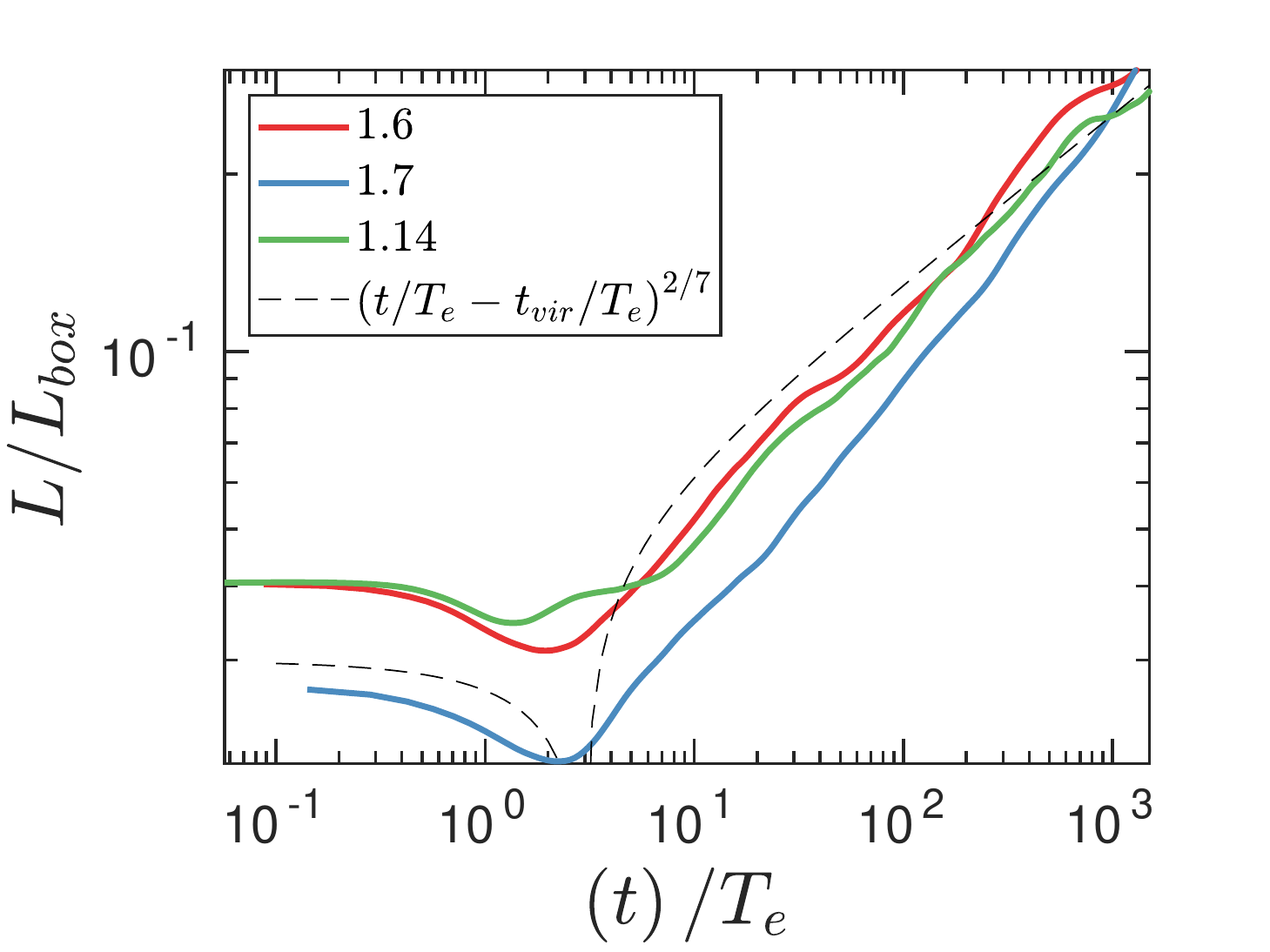}
    \includegraphics[clip,trim=00 0 0 0,width=0.5\textwidth]{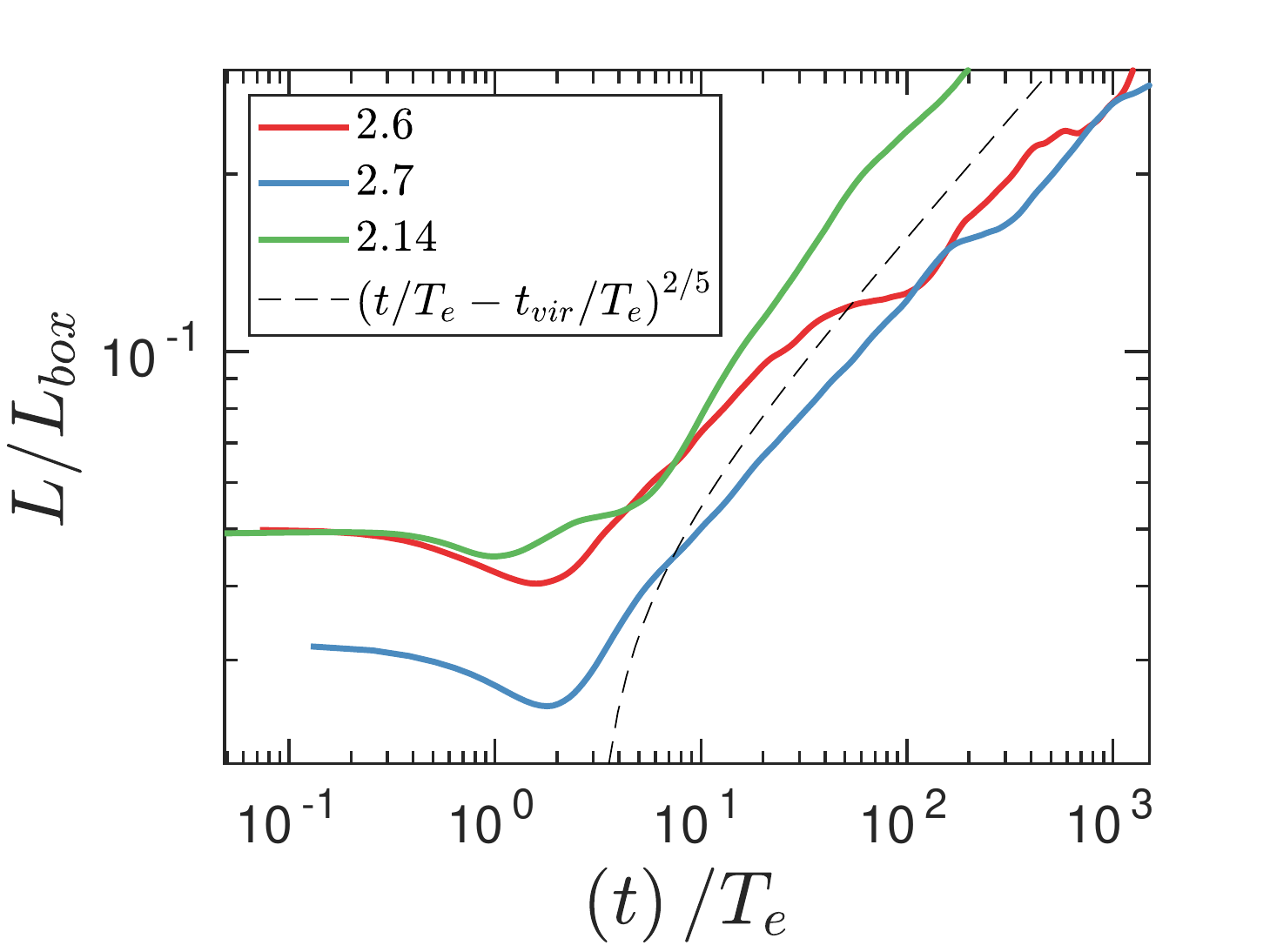}
\begin{picture}(0,0)
 \put(0,150){\small {(a)} }
\put(195,150){\small {(b)} }
   \end{picture}
\caption{\label{fig:len} Scaling of integral length with time for (a) $\kappa^{4}$ and (b) $\kappa^{2}$; $t_{vir}/T_{e}= 3$, but other nearby choices for it do not make a huge difference to our conclusions.
 }
\end{figure}

\begin{figure}
          \includegraphics[clip,trim=00 0 0 0,width=0.5\textwidth]{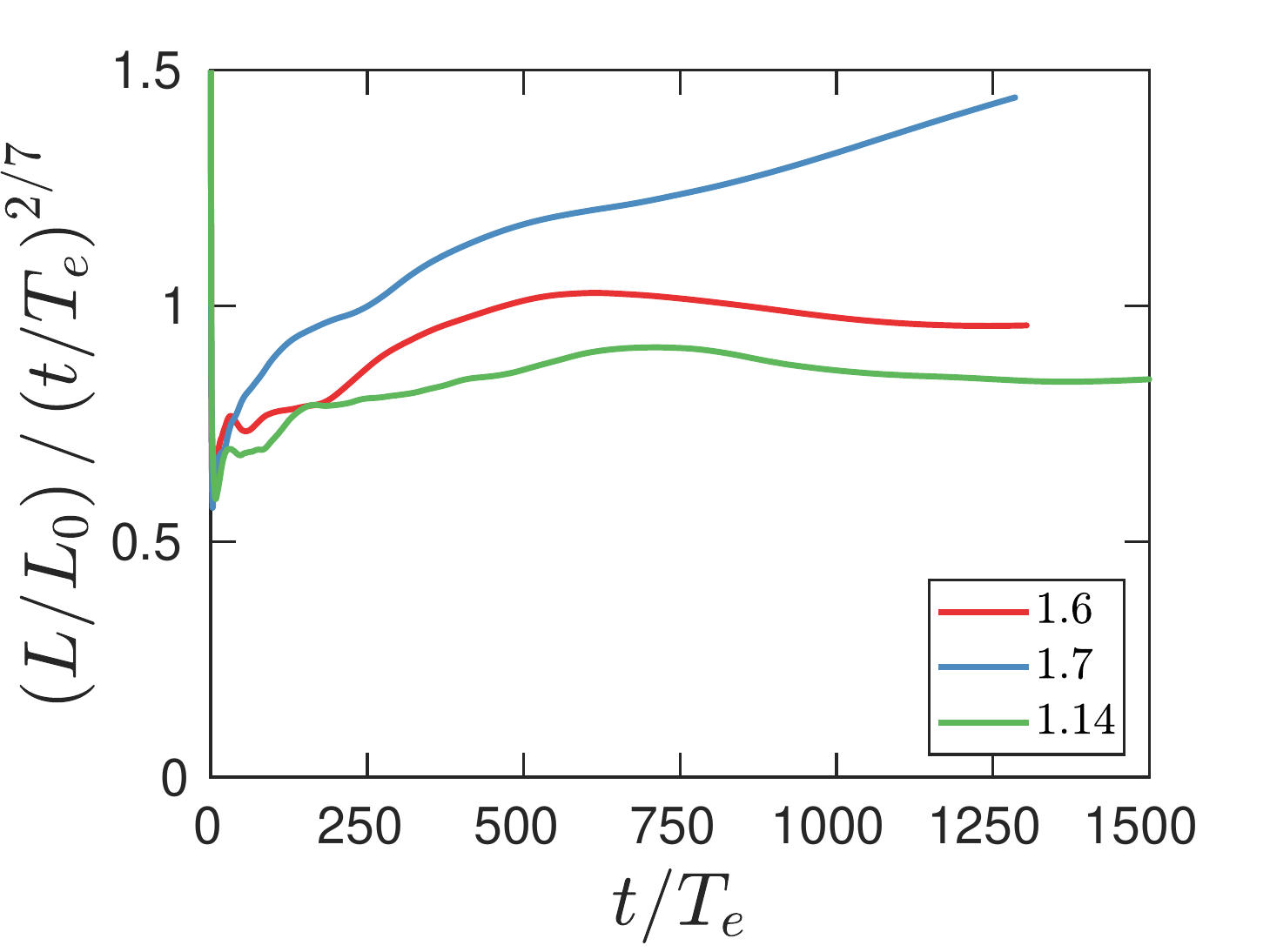}
  \includegraphics[clip,trim=00 0 0 0,width=0.5\textwidth]{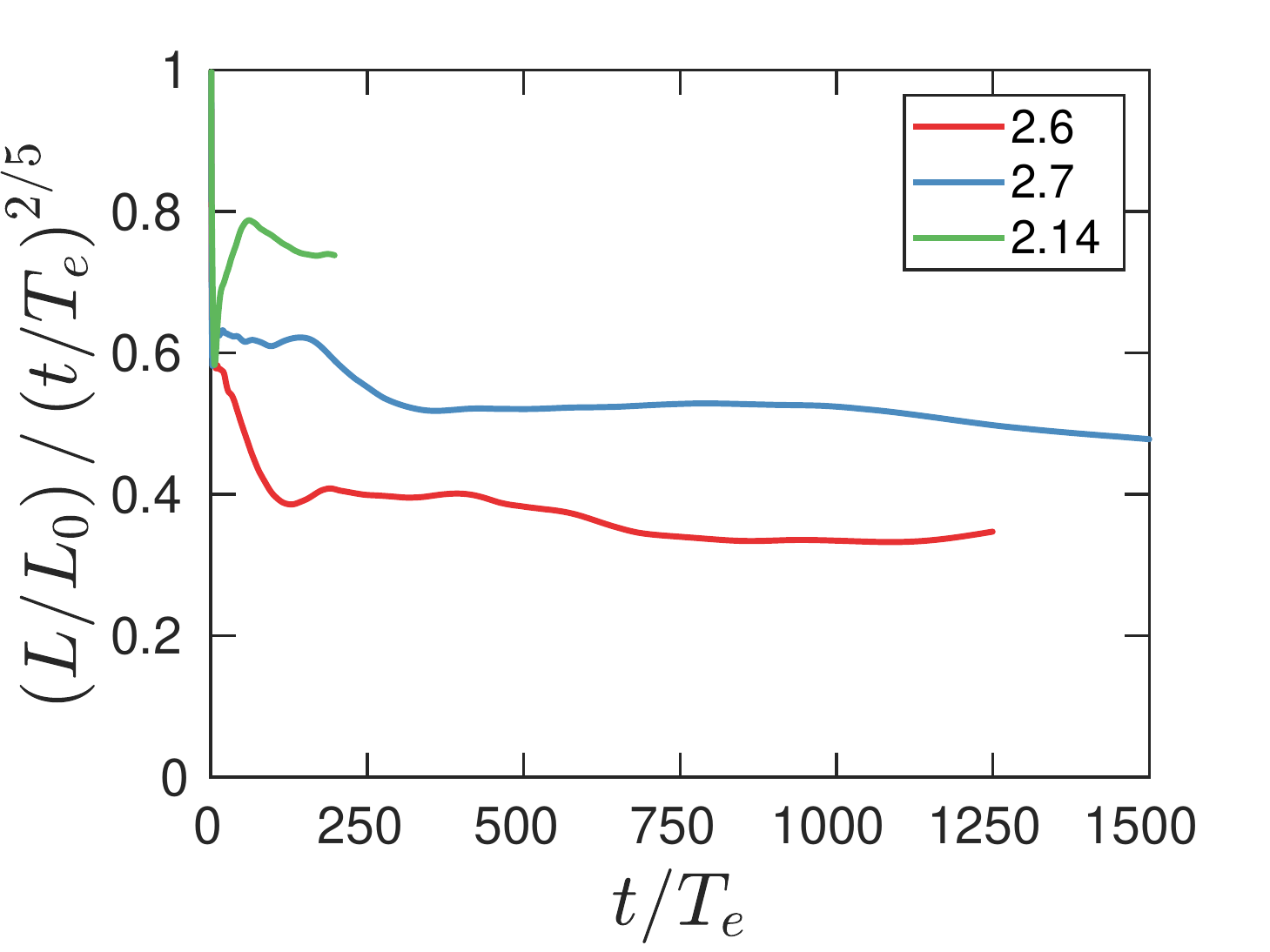}
  \begin{picture}(0,0)
 \put(0,150){\small {(a)} }
\put(195,150){\small {(b)} }
   \end{picture}
\caption{\label{fig:normlen} Normalized integral length scale for best cases with (a) $k^{4}$ and (b) $k^{2}$ spectra at the origin initially. 
 }
\end{figure}
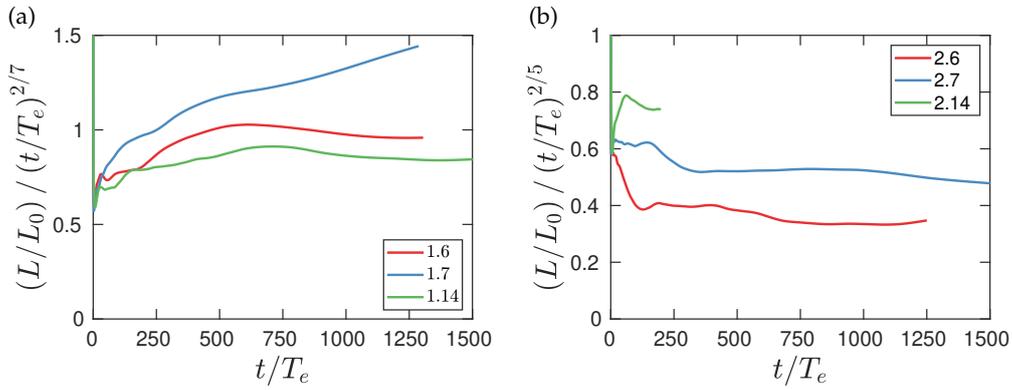

The decay process eliminates small scales preferentially so that the integral scales grow. It follows from (2.2) that the expected behaviors are
\begin{equation}
    L \sim t^{2/7} \hspace{2cm}  L \sim t^{2/5}.
\end{equation}
for Kolmogorov and the Birkhoff-Saffman cases, respectively. We evaluate the length scales using the formula stated below (1.3) and examine their behaviours in Figs.\ \ref{fig:len} and \ref{fig:normlen}, in log-log form and compensated forms, respectively. One can improve the agreement in the log-log plot between the theory (shown by the dashed line) and measurements by varying the virtual origin, but we simply show the effect of using one with three initial integral time scales. While the agreement with expectations is not perfect in either case, it appears to be reasonable.

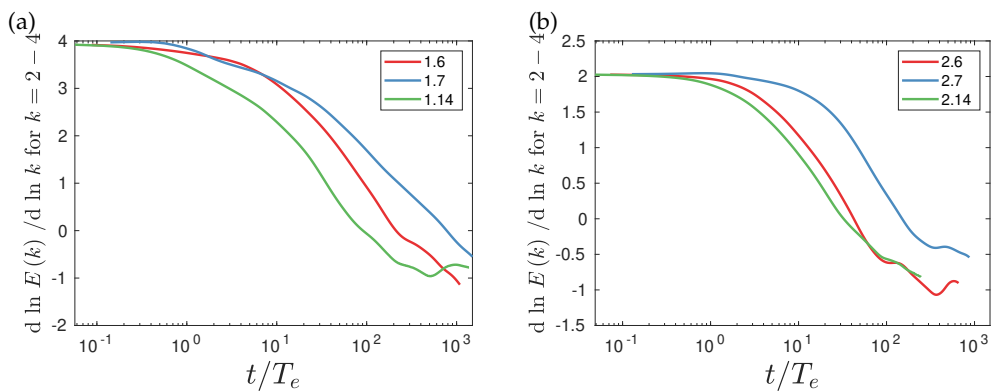
\begin{figure}
           \includegraphics[clip,trim=00 0 0 0,width=0.5\textwidth]{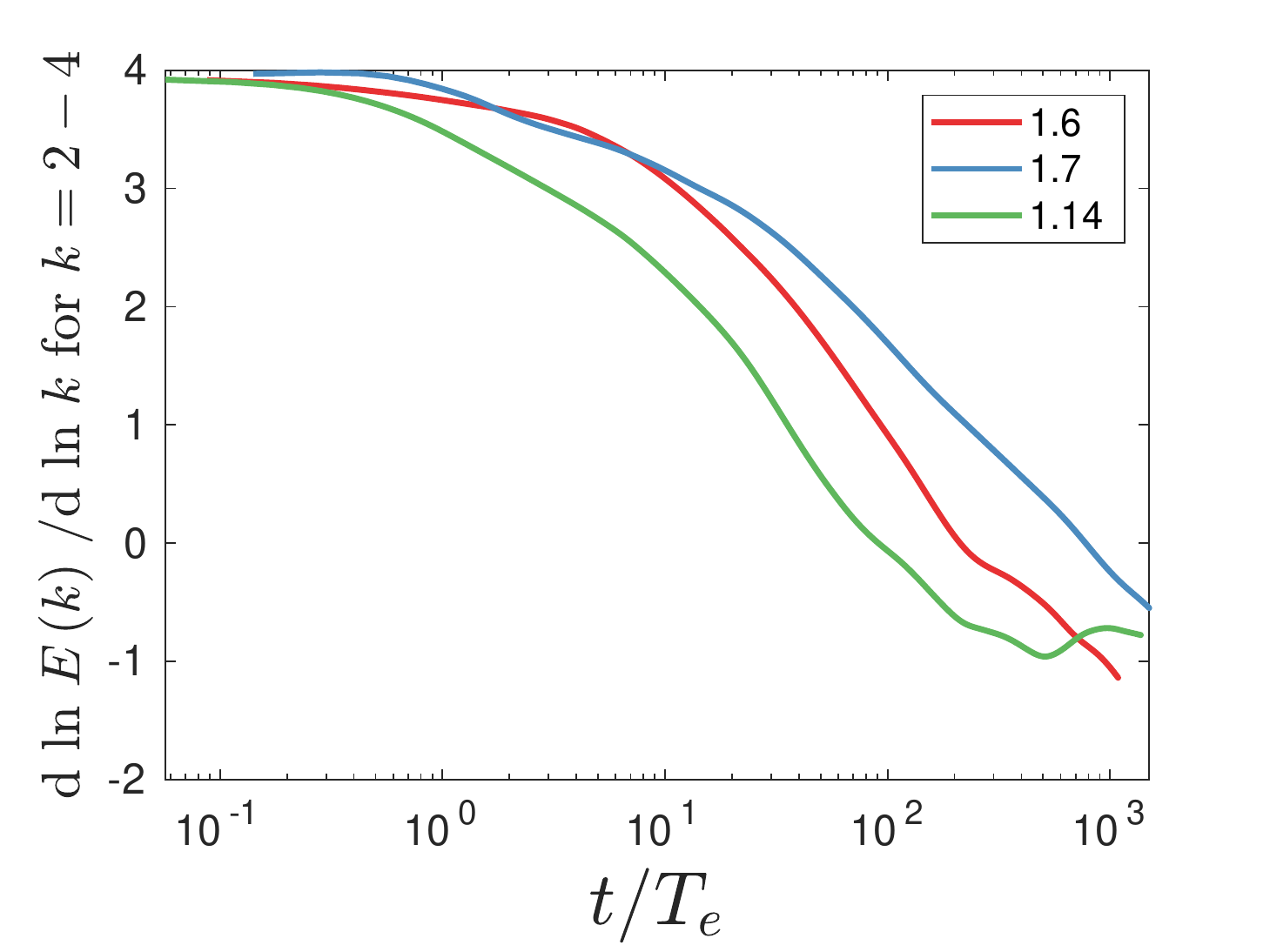}
   \includegraphics[clip,trim=00 0 0 0,width=0.5\textwidth]{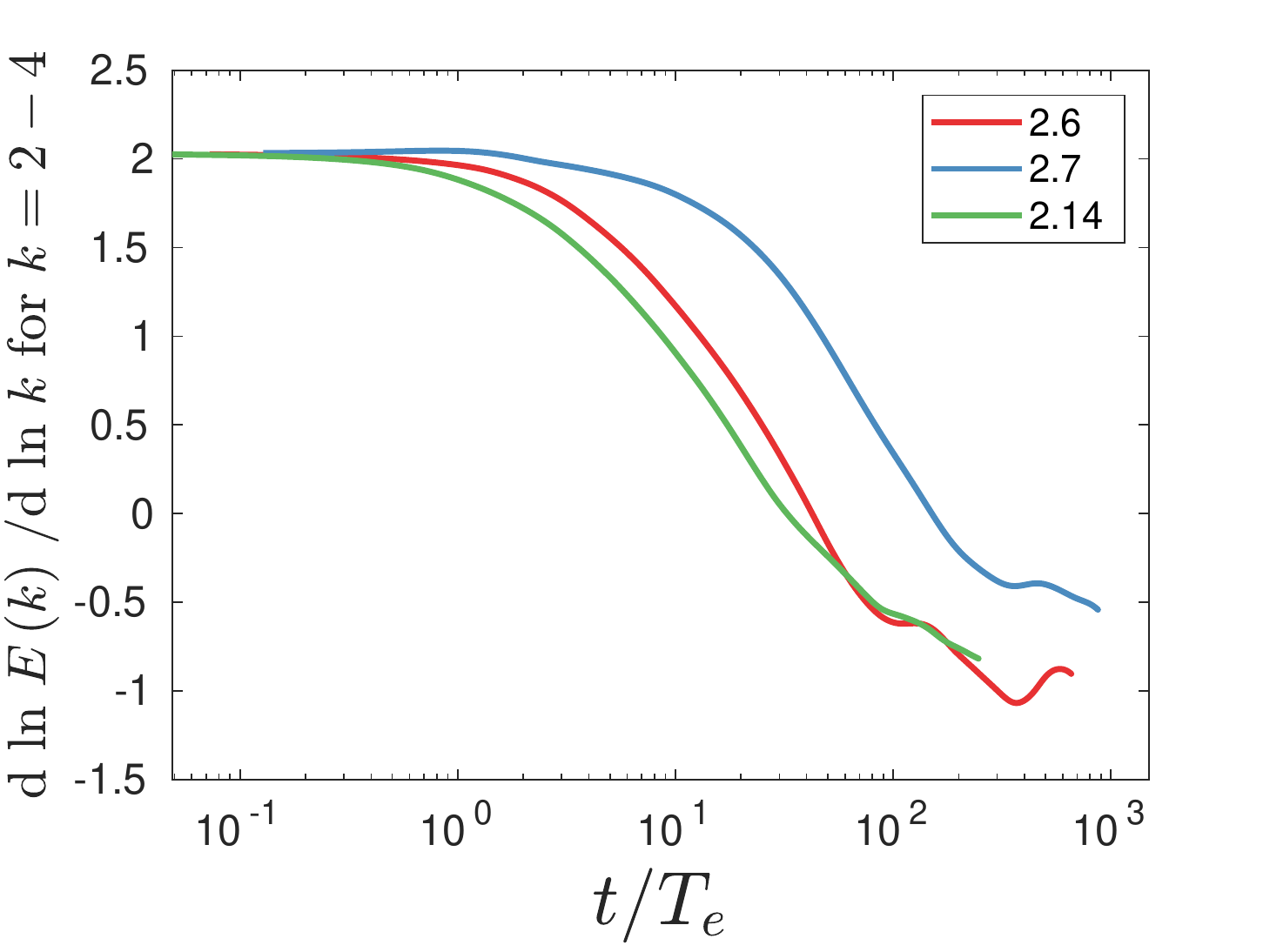}
\begin{picture}(0,0)
 \put(0,150){\small {(a)} }
\put(195,150){\small {(b)} }
   \end{picture}
\caption{\label{fig:permextra} Evolution of the exponent $q$ of the energy spectrum of the form $E \left(\kappa \right) \propto \kappa ^{q}$ at low wavenumbers computed using just the first three wavenumbers for the cases chosen. A constant with time indicates permanence of large eddies. (a) $E\left( \kappa \right) \propto \kappa^{4}$ and (b) $E\left( \kappa \right) \propto \kappa^{2}$. Only for short periods of time is the slope close to 4 in (a) and to 2 in (b). The erosion sets in relatively quickly in all simulations of table 1.
 }
\end{figure}

\section{Discussion and Conclusions}
We have studied the $\kappa^{4}$ and the $\kappa^{2}$ cases using well-resolved DNS for homogeneous and isotropic turbulence (HIT) for  a variety of conditions (table 1). Our results show that for an unambiguous power law for energy decay to exist, one needs to start with high enough Reynolds number and the energy-containing wavenumber
is far enough away from dissipation scales, while allowing a power law at the origin. Only a few cases we examined satisfy these criteria. These cases also obey the requirement on the skewness of the velocity gradient. We found that when the power-law at the origin was initiated with a $\kappa^4$-type behavior, the Kolmogorov decay law with the exponent close to $10/7$ was indeed observed. The $\kappa^2$ turbulence conforms just as well to its own expectations.  

At the end of this exercise, it is somewhat disappointing that the results are not more closely aligned with theoretical arguments. We believe that this is due to the fact that the $\kappa^4$ and $\kappa^2$ behaviors at the origin do not persist for long periods of time; at least for the conditions of the present simulations, nonlinear interactions diminish their extent rather rapidly; see Fig.~\ref{fig:permextra} which plots the slope neat the origin as a function of time for the cases examined. This suggests that in neither case do the invariants exist exactly except approximately for an initial period of time. Indeed, as argued by Yakhot \cite{yakhot}, that approximate condition seems to be adequate for the Kolmogorov decay law; one presumes likewise for the $\kappa^2$ behavior. Overall, it appears that for a power-law decay in energy to exist according to one of the theories, one should set off the initial conditions correctly even if one may not require strict permanence of the large scale through out the decay. This appears true for exponents for the energy decay and the length scale growth.

While considering the decay problem recently, Mccomb \cite{McPF2016} made a number of interesting theoretical points. One of them is that the Birkhoff-Saffman invariant is strictly zero for HIT. He also pointed out that this does not rule out that some such entity may exist for computational turbulence in a finite box, which is essentially what we have found. The somewhat unsatisfactory results we have found show that it is hard to satisfy all the conditions exactly, even in box-type HIT, but our demonstrations suggest that one can approach them reasonably closely. It is not clear to us, after our experience with the many simulations listed in table 1, that permanence of large scales can be established in any finite system.

Finally, it is worth commenting on the diversity of exponents observed in experiments and simulations (Fig.\ 1). Indeed, if one did not pay any attention to maintaining a particular spectral behavior near the origin, a range of decay exponents were observed in simulations (often there were no good power laws). However, some correspondence with the theoretical expectations were obtained when we started the spectrum at the origin in a certain way. There is experimental evidence to suggest that initial conditions set off the decay with certain characteristics \cite{sreeni1984}. We speculate that the experimental conditions are influenced by several of these issues. Besides, the practice of choosing arbitrary virtual origins to fit the limited range of decay data by power-laws, and of choosing to fit distances as close to the grids of 5 mesh lengths, could result in a plethora of uncertain results. We think that Fig.\ 1 reflects these uncertainties.

\vspace{0.5cm}

Acknowledgments: We are pleased to present this short paper in the collection honoring the remarkably long and productive scientific career of Professor Uriel Frisch. We acknowledge support from the National Science Foundation (Grant No. 1605914) and from the Extreme Science and Engineering Discovery Environment (XSEDE) for computational resources.

\bibliographystyle{RS}
\bibliography{all.bib}

\end{document}